# Temperature-Dependent Spectroscopy of $Cr^{3+}$:YGG Nanophosphors with Multisite Emission

Mykhailo Chaika

Institute of Low Temperature and Structure Research Polish Academy of Science, ul. Okolna 2, 50-422, Wroclaw, Poland

## Abstract

An important feature of $Cr^{3+}$ is the ability to tune the absorption and emission spectra by changing the host. However, in some cases, different emission spectra can be detected in samples with similar Racah parameters. The present paper reports changes in the luminescence properties of $Cr^{3+}$:YGG nanocrystals. $Cr^{3+}$:YGG nanocrystals were synthesized by a modified Pechini method to obtain nanocrystals containing two fractions of $Cr^{3+}$:YGG nanocrystals with different lattice parameters and crystalline sizes. The increase in the concentration of $Cr^{3+}$ ions leads to a redshift of the $^4T_{2g}$ emission from 750 nm to 850 nm for luminescence spectra measured at 80K. In contrast, room temperature luminescence spectra showed similarity in the shape of the emission spectra. The calculated crystal field strength was found to increase with the concentration of $Cr^{3+}$ ions, so the detected patterns in low-temperature luminescence spectra were explained by energy transfer chain processes.

## 1. Introduction

$Cr^{3+}$-doped phosphors are widely used for their tunable luminescence, but performance variations highlight the need for further investigation into their spectroscopic behavior. Cr-doped phosphors play an important role in daily life. $Cr^{3+}$-doped garnets are one of the best light converters for NIR LEDs [1,2], $Cr^{3+}$:$Al_2O_3$ was the first solid-state laser [3,4] and continues to play an important role today, and $Cr^{3+}$-doped phosphors are also often used for luminescent thermometry [5]. This wide range of applications of $Cr^{3+}$-doped materials is due to their ability to tune the luminescence properties by changing the hosts [6]. For example, the room temperature emission spectra gradually change from a sharp and narrow red emission for $Cr^{3+}$:$Al_2O_3$ to a broadband NIR emission for $Cr^{3+}$:$ZnWO_4$ [7]. All these applications rely on the spectroscopic characteristics of $Cr^{3+}$ ions, such as the spectral position and emission intensity [8,9]. One of the challenges encountered in this context is the variability in the properties of Cr-doped phosphors, even for the same chemical composition [10]. Investigating these irregularities will help to improve their functional characteristics.

The luminescence properties of $Cr^{3+}$ ions are influenced by the interaction between a $Cr^{3+}$ ion and an octahedral crystal field, which modifies energy transitions and spectral properties. $Cr^{3+}$ ions belong to the $3d^3$ family of transition metal ions and have a simple electronic structure with a $^4F$ ground state and lowest excited-state free ion energy levels $^4P$, $^2G$, and $^2F$ [7]. The interaction between a $Cr^{3+}$ ion and an octahedral crystal filed octahedral crystal field causes a splitting of the free ion energy levels. The $^4F$ free ion ground state splits into a $^4A_{2g}$ ground level and two excited levels, $^4T_{2g}$ and $^4T_{1g}$ [11]. Similarly, the $^2G$ splits into $^2T_{2g}$, $^2T_{1g}$, and $^2E_g$ energy levels [12]. The transition between the $^4A_{2g}$ ground level and the $^4T_{2g}$ and $^4T_{1g}$ levels is strongly dependent on the crystal field strength [13]. In contrast, the transitions between the $^4A_{2g}$ ground level and the $^2Eg$, $^2T_{1g}$, and $^2T_{2g}$ levels are almost independent of the crystal field strength. Therefore, a change in the crystal field affects the positions of the $^4A_{2g}\leftrightarrow{}^4T_{2g}$ and $^4A_{2g}\leftrightarrow{}^4T_{1g}$ electronic transitions, while the $^4A_{2g}\leftrightarrow{}^2E_g$ transitions remain unchanged. This causes changes in both the absorption and emission spectra for different values of crystal field strength [14].

Depending on the type of lower excited state, $Cr^{3+}$-doped phosphor at low temperatures might exhibit narrow R-lines or broadband NIR emission. The spectrum of $Cr^{3+}$-doped phosphors can be divided into two categories characterized by strong and weak crystal fields. This deviation is based on the position of two lower levels: $^4T_{2g}$ and $^2E_g$. When measured under liquid helium temperatures, samples where $Cr^{3+}$ ions have the lowest excited energy level $^4T_{2g}$ (weak crystal field) are characterized by only broadband emission ($^4T_{2g}\rightarrow{}^4A_{2g}$). Consequently, samples where the lowest excited energy level is $^2E_g$ (strong crystal field) are characterized by narrow R-lines emission ($^2E_g\rightarrow{}^4A_{2g}$) [12]. For ions in a strong crystal field, an increase in temperature increases the possibility of populating the higher $^4T_{2g}$ energy level, so both R-lines ($^2E_g$) and broadband ($^4T_{2g}$) emission can be detected simultaneously. In contrast, $Cr^{3+}$ ions in a low crystal field exhibit only emission from the $^4T_{2g}$ energy level [7]. In some cases, when the $^4T_{2g}$ energy level is slightly above or below the $^2E_g$ energy level, both broadband and R-lines can be detected simultaneously, even at 4K [14]. The presence of R-lines in the $Cr^{3+}$ emission spectra clearly indicates that the emission originates from ions in a strong or intermediate crystal field. In contrast, the presence of broadband emission with $\lambda_{max}$>800 nm indicates the presence of $Cr^{3+}$ ions in a low crystal field [14].

The difference in the emission spectra might also be explained by the energy transfer between the $Cr^{3+}$ ions in various local environments, previously named as "chain" of energy transfer [10]. Spectroscopic properties of $Cr^{3+}$:GGG nanoceramics showed a change in the shape of the low-temperature emission spectra with the change in $Cr^{3+}$ concentration [10]. An increase in $Cr^{3+}$ concentration caused a proportional increase in the ratio of broadband $^4T_{2g}(^4F)$ emission intensity to the overall $Cr^{3+}$ emission. The difference in the measured spectra was unexpected since the calculated Racah parameters were in proximity for the measured samples. Moreover, the measured room-temperature emission spectra were almost identical, confirming the similarities in the Racah parameters. Additionally, there were no discernible differences in the peak position and width of the $^4T_{2g}(4F)\rightarrow{}^4A_{2g}(4F)$ broadband emission across the entire temperature range [10]. The variation in low-temperature emission spectra was explained by energy transfer between chromium ions in sites with different local crystal field strengths. It was shown that the nanoceramics contained multiple $(CrO_6)^{9-}$ optically active centers ($Cr^{3+}$ ions in octahedral sites) with variations in crystal field strength, as well as $Cr^{3+}$ in a low crystal field. It was proposed that $Cr^{3+}$ ions create a "chain" of energy transfer between themselves, but the presence of a $Cr^{3+}$ ion in a low crystal field breaks this "chain" of energy transfer [10]. $Cr^{3+}$ ions in low crystal field sites demonstrated a higher probability of emitting photons rather than transferring excitation energy to nearby $Cr^{3+}$ ions. This energy transfer increased with the concentration of $Cr^{3+}$ ions, consequently elevating the ratio of the emission intensity of $Cr^{3+}$ ions in a low crystal field to the total luminescence.

The earlier study was based on samples with a low concentration of $Cr^{3+}$ ions, which makes it difficult to confirm the energy transfer chain theory. The $Cr^{3+}$ ion concentration varied in a small range from 0.1 at.% to 0.3 at.% for $Cr^{3+}$:GGG nanoceramics [10], which introduces some uncertainty in the proposed explanation. It cannot be ruled out that the detected changes to the luminescence spectra are due to a decrease in the crystal field strength, which would disprove the proposed energy transfer chain theory. However, the possible presence of a chain energy transfer phenomenon in Cr-doped phosphors would result in changes to the spectroscopic properties. For example, it could affect the luminescence intensity ratio in luminescence thermometers or lead to an increase in optical loss in laser materials. Confirming the existence of the chain energy transfer phenomenon in $Cr^{3+}$-doped phosphors requires a complex study.

This paper focuses on the study of the spectroscopic properties of $Cr^{3+}$:YGG nanopowders, focusing on energy transfer between ions in different crystal field environments. The concentration of $Cr^{3+}$ ions varied from 0 to 30 at.%. Excitation, emission, and luminescence decay curves were collected in the temperature range of -190 °C to 500 °C. It was shown that increasing the concentration of $Cr^{3+}$ ions caused a redshift in the emission spectra. The detected features were explained by the energy transfer process between $Cr^{3+}$ ions in octahedral sites with different local crystal field strengths.

## 2. Experimental

Cr:YGG nanocrystals were synthesized by the Pechini method with the following compositions: $Y_3Ga_{(1-x)2}Cr_{2x}Ga_3O_{12}$, where x = 0, 0.0025, 0.0125, 0.025, 0.075, 0.125, 0.25, and 0.75, respectively. The obtained $Cr^{3+}$ concentrations were 0.1 at.% ($2.2 \cdot 10^{19}$ $cm^{-3}$), 0.5 ($1.1 \cdot 10^{20}$ $cm^{-3}$), at.%, 1 at.% ($2.2 \cdot 10^{20}$ $cm^{-3}$), 3 at.% ($6.5 \cdot 10^{20}$ $cm^{-3}$), 5 at.% ($1.1 \cdot 10^{21}$ $cm^{-3}$), 10 at.% ($2.2 \cdot 10^{21}$ $cm^{-3}$), and 30 at.% ($6.5 \cdot 10^{21}$ $cm^{-3}$), relative to $Ga^{3+}$ ions. The following reagents were used as starting compounds: yttrium oxide ($Y_2O_3$, Stanford Materials Company – 99.999% purity), chromium nitrate hydrate ($Cr(NO_3)_3 \times 9H_2O$, – 99.999% purity), gallium chloride ($Cl_3Ga$, Thermo Scientific Chemicals – 99.999% purity), citric acid nonhydrate (C6H8O7, POCH – 99.5% purity), and ethylene glycol ($C_2H_6O_2$, POCH – 99% purity). Initially, stoichiometric amounts of yttrium oxides and gallium chloride were converted into nitrates. Then, aqueous nitrate solutions were mixed for 3 hours with non-aqueous citric acid and ethylene glycol at a molar ratio of 8:5 relative to the final product. The samples were then placed in a drying chamber at 90 °C for seven days; as a result, a brown resin was obtained. Finally, the resin was calcined at 900 °C in air for 16 hours. The sintering conditions of the samples were chosen to synthesize a double garnet structure.

Structural and optical characterization of the Cr:YGG nanocrystals was performed using XRD, TEM, absorbance, and photoluminescence spectroscopy. The X-ray diffraction patterns were measured using a Panalytical X'Pert pro X-ray powder diffractometer. Rietveld refinements were performed using the FullProfSuite program with the WinPLOTR and WinPLOTR-2006 applications. The synthesized samples were analyzed via transmission electron microscopy (TEM) using a Philips CM-20 SuperTwin microscope. Absorbance spectra were measured using a Varian 5E UV–VIS-NIR spectrophotometer. Excitation and emission spectra, as well as luminescence lifetimes, were measured using a FLS1000 fluorescence spectrometer. A 450 W ozone-free xenon lamp and a xenon flash lamp were used as excitation sources.

## 3. Result

### *3.1 Microstructure*

XRD analysis showed that the main crystalline phase in Cr:YGG samples is garnet, with impurity phases appearing at higher $Cr^{3+}$ concentrations. The microstructure of the synthesized Cr:YGG nanocrystals was analyzed by powder X-ray diffraction (XRD). The main peaks on the collected diffractograms correspond to the garnet phase with an Ia-3d space group and a chemical composition of $Y_3Ga_{(1-x)2}Cr_{2x}Ga_3O_{12}$ [15,16]. The XRD patterns of the Cr10 and Cr30 samples showed the presence of two impurity phases: $Y_2O_3$ (Ia-3 space group, cell parameter a - 12.293(2) Å) and $Ga_2O_3$ (C2/m space group, cell parameters a - 12.222(7) Å, b - - 3.039(2) Å, c - 5.806(3) Å, β - 103.7626(1) °. The concentration of $Y_2O_3$ was 1.1(1) wt.%(Cr10) and 11(1) wt.% (Cr30), while that of $Ga_2O_3$ was 2.6(2) wt.% (Cr10), and 6.0(5) wt.% (Cr30).

XRD analysis shows no impurity phases in the Cr0 – Cr5 samples. However, both $Y_2O_3$ and $Ga_2O_3$ phases might still be present in all samples. XRD can confirm phase purity only up to a certain sensitivity threshold. Due to peak overlap with the garnet phase, it is difficult to confirm the presence or absence of $Ga_2O_3$ and $Y_2O_3$ below 0.5 wt.%. Earlier studies on Cr:YAG transparent ceramics have shown that minor concentrations of sesquioxides are always present in the samples [17]. Based on previous reports, a small amount of $Cr^{3+}$:$Ga_2O_3$ in the sample may be difficult to distinguish from the $Cr^{3+}$:YGG [18].

Table 1: The result of the Rietveld analysis: $R_{Brag}$ - Bragg R-factor, $R_f$ - Rf-factor, d – crystalline size.

| Sample | Phase | $R_{Brag}$ (%) | $R_F$ (%) | Lattice (A) | d (nm) | Weight (%) |
|---|---|---|---|---|---|---|
| Cr0 | YGG(1) | 2.66 | 3.29 | 12.296(1). | 31(4) | 19.18( 3.10) |
| | YGG(2) | 2.13 | 3.03 | 12.350(3) | 16(1) | 80.82( 4.42) |
| Cr01 | YGG(1) | 2.57 | 3.18 | 12.296(1). | 50(5) | 11.71( 1.37) |
| | YGG(2) | 3.14 | 3.93 | 12.346(2) | 19(2) | 88.29( 2.56) |
| Cr05 | YGG(1) | 2.29 | 3.09 | 12.289(7). | 51(1) | 23.21( 1.44) |
| | YGG(2) | 3.24 | 3.94 | 12.338(2) | 19(2) | 76.79( 2.34) |
| Cr1 | YGG(1) | 2.16 | 3.26 | 12.2846(8). | 42(2) | 53.12( 3.56) |
| | YGG(2) | 2.58 | 3.26 | 12.357(2) | 21(2) | 46.88( 3.45) |
| Cr3 | YGG(1) | 2.03 | 3.31 | 12.281(1). | 35(2) | 52.29( 2.92) |
| | YGG(2) | 3.03 | 2.78 | 12.357(3) | 21(2) | 47.71( 2.62) |
| Cr5 | YGG(1) | 2.32 | 3.16 | 12.290(1). | 31(2) | 58.81( 2.94) |
| | YGG(2) | 3.46 | 3.59 | 12.383(3) | 20(2) | 41.19( 2.43) |
| Cr10 | YGG(1) | 2.24 | 2.59 | 12.2871(8). | 47(2) | 28.19( 2.19) |
| | YGG(2) | 2.33 | 2.65 | 12.342(2) | 19(2) | 68.10( 3.02) |
| | $Y_2O_3$ | 4.62 | 4.05 | 12.293(2) | - | 1.13( 0.08) |
| | $Ga_2O_3$ | 8.13 | 4.97 | 12.222(7)<br>3.039(2)<br>5.806(3)<br>103.7626(1) ° | - | 2.58( 0.16) |
| Cr30 | YGG(1) | 3.25 | 3.01 | 12.282(1). | 51(1) | 57.13( 7.42) |
| | YGG(2) | 3.43 | 3.21 | 12.292(2) | 40(5) | 25.61( 5.77) |
| | $Y_2O_3$ | 8.98 | 6.67 | 12.293(2) | - | 11.22( 0.95) |
| | $Ga_2O_3$ | 23.8 | 11.9 | 12.222(7)<br>3.039(2)<br>5.806(3)<br>103.7626(1) ° | - | 6.04( 0.54) |

The X-ray diffraction (XRD) patterns were analyzed by Rietveld refinement using the FullProf suite. Two garnet phases were used because a single-phase mode did not give a satisfactory fit. The background was resolved using a polynomial of 9th degree. The lattice parameters for both garnets were refined. Sample displacement and zero shift were refined iteratively. The diffraction peak shape was described by a pseudo-Voigt profile, and the Cagliotti half-width parameters U, V, W were refined for two phases separately. All atom positions that could be refined were allowed to vary. Individual atomic displacement parameters (ADPs) were constrained to a single common isotropic value, which was refined. Profile asymmetry parameters were also refined.

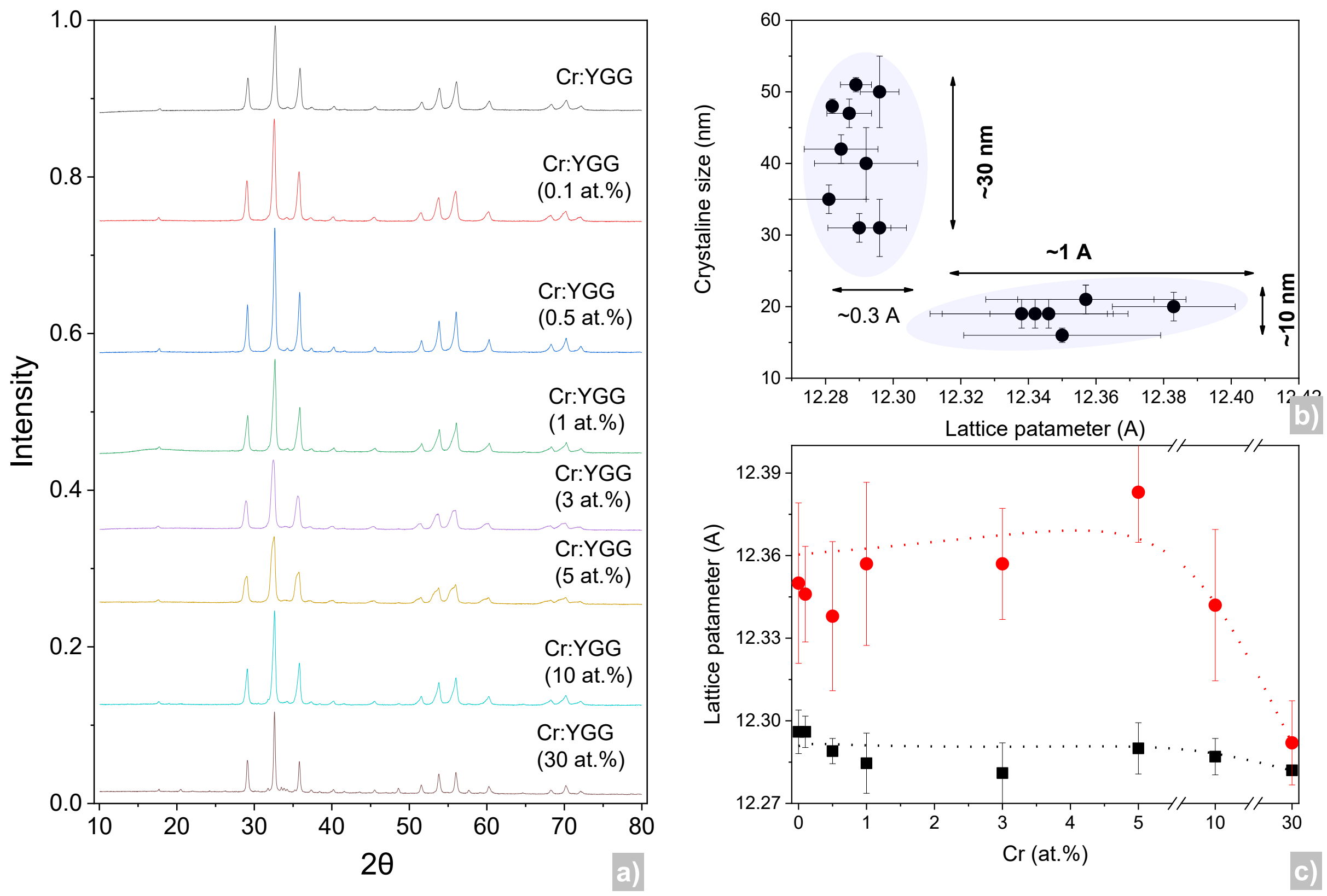

Fig. 1:a) XRD pattern of the synthesized Cr:YGG nanocrystals; The results of the Rietveld analysis b) lattice parameter and average crystalline size, c) dependence of the lattice parameters of both garnet fractions on the concentration of $Cr^{3+}$ ions.

The shape and width of the garnet peaks indicate the presence of two fractions of the garnet phase with different lattice parameters. The collected lattice parameters and the average size of the respective crystallites are presented in Tables 1, S1, and Fig. 1(b). The strong overlapping of the garnet peaks makes it difficult to calculate the lattice parameters with high accuracy. Fig. 1(b) shows the change in the lattice parameters and average crystalline size of different Cr:GGG nanocrystals based on the Rietveld analysis (Fig. S1). The presence of two fractions is clearly visible: the larger ones, with an average crystalline size of 30-50 nm, and the smaller ones 16-21 nm. The crystalline size was calculated by the Scherrer equation. The larger fraction is characterized by a relatively small lattice parameter (12.284-12.302 Å), while the smaller fraction has a larger lattice parameter (12.301-12.408 Å). The presence of two garnet phases with different microstructures allows the redistribution of $Cr^{3+}$ ions between the octahedral sites with different local crystal field strengths.

The change in $Cr^{3+}$ ion concentration changes the distribution of garnet among the smaller and larger fractions. This influence can be observed by monitoring the Full Width at Half Maximum ($\Delta_{FWHM}$) of the most intense garnet peak (2θ-32.5 °). The measured $\Delta_{FWHM}$ values for Cr0, Cr01, Cr05, Cr1, Cr3, Cr5, and Cr10, were 0.42 °, 0.42 °, 0.33 °, 0.42 °, 0.57 °, 0.63 ° 0.39 °, and 0.253 °, respectively. Samples with 1 at.% $Cr^{3+}$ ion concentration or lower exhibited narrower peaks compared to Cr3 and Cr5 samples. In Cr3 and Cr5 samples, the different garnet peaks became distinguishable (Fig. S2). These changes are attributed to variations in the ratio between smaller and larger fractions. For Cr0, Cr,01, Cr05, and Cr1 samples, the microstructure consisted of roughly ¼ of larger fractions and ¾ of smaller fractions. In contrast, Cr3 and Cr5 samples exhibited a roughly equal ratio of smaller and larger fractions. Further increasing the $Cr^{3+}$ concentration disrupted the formation of the garnet phase, leading to the appearance of $Y_2O_3$ and $Ga_2O_3$ impurity phases.

TEM analysis revealed the nanoparticles were ~50 nm in size, and consisted of crystalline, approximately 10 nm in size. The microstructure of the samples was investigated using TEM (Fig. 2). The samples consist of nanoparticles with an average size of approximately 50 nm. They are characterized by a significant contrast difference in the individual grain surface, showing "darker areas" that are ~10 nm in thickness. These dark areas might be associated with the polycrystalline nature of the synthesized powders. The intensity difference indicates variation in the electron density, which could be due to the difference in thickness or in orientation of the individual grains. The TEM images suggest that these nanoparticles have a round shape without any spikes, which could explain the variations in particle thickness on the scale of ~10 nm. Therefore, the contrast observed is likely due to the difference in the orientation of the grains. It should be noted that the images of our samples differ from previously observed TEM images of "monofractional" Cr:YAG [19], and Cr,Yb:YAG [20] nanocrystals. This observation is consistent with the results of XRD analysis (Fig. 1(b)) and supports the conclusion that our nanoparticles have a polycrystalline nature, containing both larger grains (~50 nm) and small grains (~ 10 nm).

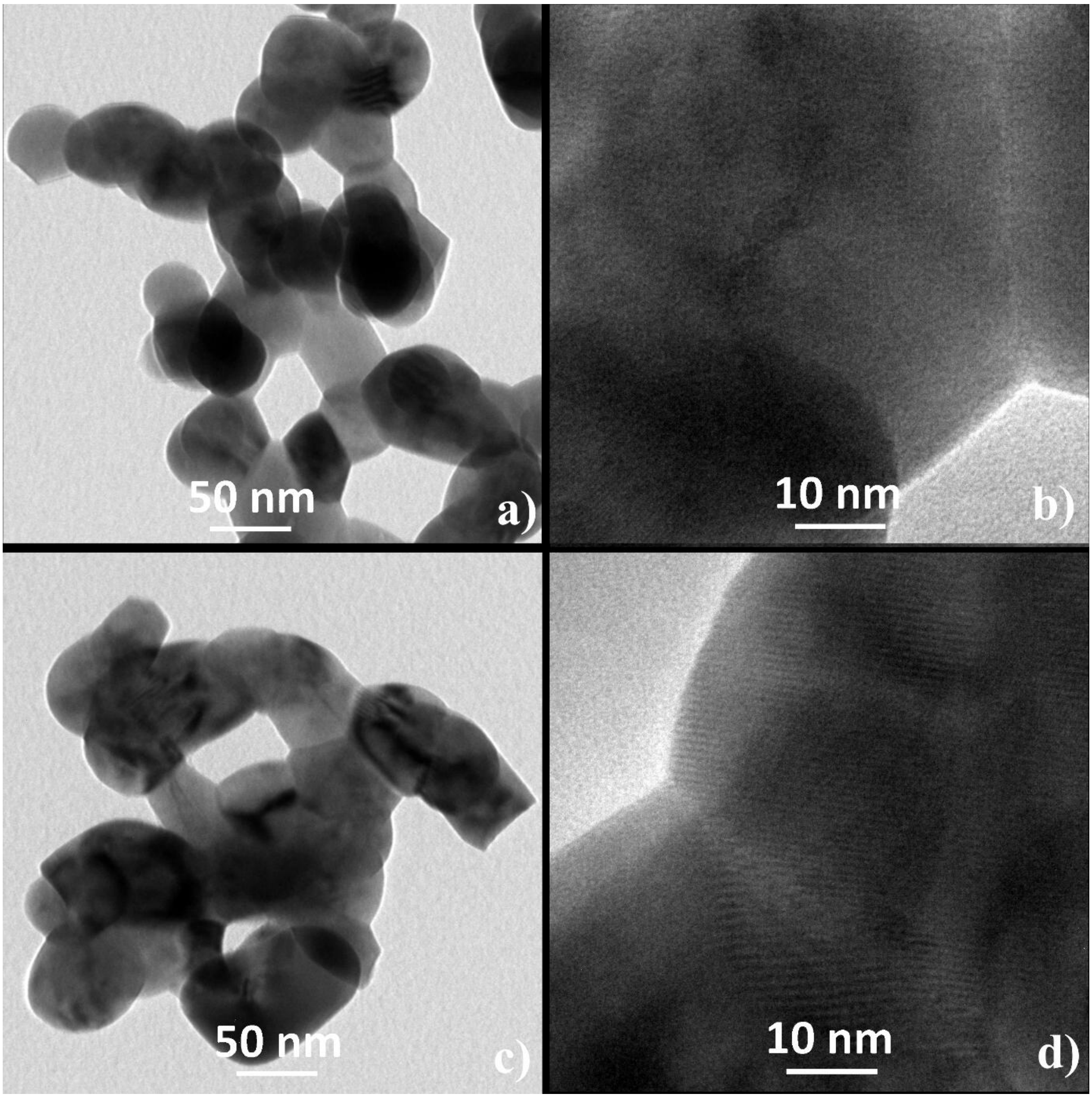

Fig. 2: TEM image of $Cr^{3+}$:YGG nanocrystals a),b) (0,1 at.%), c),d) (3 at.%)

*3.2 Diffuse reflectance spectra*

Diffuse reflectance spectra reveal the presence of $Cr^{3+}$, color centers, and possible $Cr^{6+}$ absorption bands, together with the similarity in optical band gap. Diffuse reflectance spectra of the synthesized samples were collected in the range of 200-800 nm (Fig. 3). Four types of the optical absorption centers were detected in the samples: the first is due to the presence of $Fe^{3+}$ ions ($\lambda_{max}$ - 267 nm); the second is color centers $\lambda_{max}$ - 243 nm; the third is probably due to charge transfer from oxygen to $Cr^{6+}$ ions ($\lambda_{max}$ - 372 nm); the forth is $Cr^{3+}$ ($\lambda_{max}$ - 434 nm, 623 nm) [19,21,22].

The increase in the concentration of $Cr^{3+}$ ions caused an increase in the oxygen to $Cr^{6+}$ charge transfer absorption bands, which wasn't expected [4,19]. $Cr^{3+}$ excitation spectra showed the absence of a band at $\lambda_{max}$ - 372 nm, indicating that $Cr^{6+}$ ions (if the assumption is correct [20]) do not participate in $Cr^{3+}$ luminescence processes. Absorption edge of Cr:YGG nanocrystals decreased linearly with the increase in the concentration of $Cr^{3+}$ ions. The measured values were 5.67(1) eV (Cr0), 5.67(1) eV (Cr01), 5.65(1) eV (Cr05), 5.68(1) eV (Cr1), 5.65(1) eV (Cr3), 5.62(1) eV Cr5, 5.57(1) eV (Cr10), 5.50(1) eV (Cr30). The change in the optical band gap was insignificant and probably had little influence on the $Cr^{3+}$ luminescence.

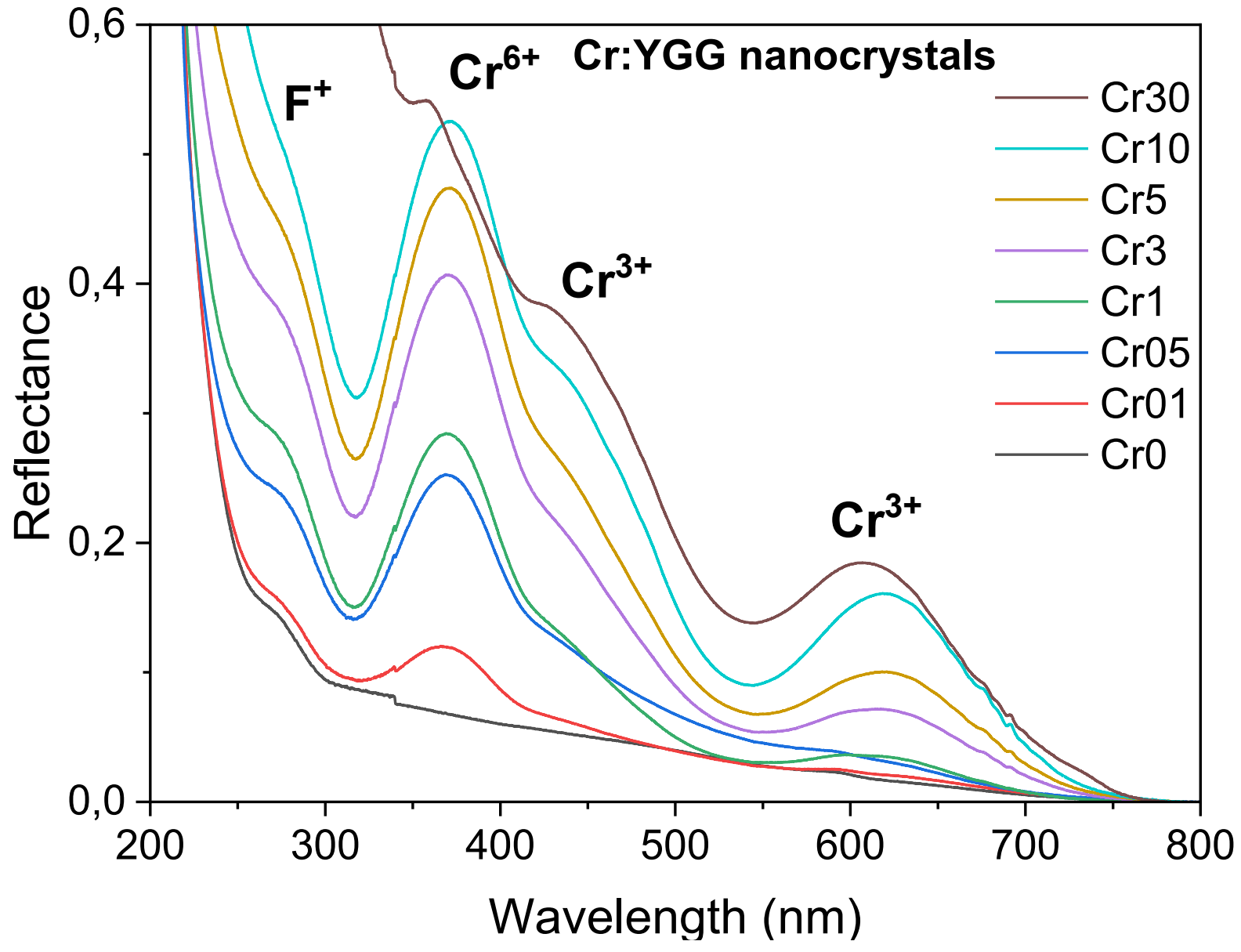

Fig. 3: Diffuse reflectance spectra of the Cr:YGG nanocrystals.

Based on calculated Racah parameters and crystal field strength, all samples are characterized by the $Cr^{3+}$ ions in a strong crystal field. Optical properties of $Cr^{3+}$ ions might be predicted based on the Racah parameters, crystal splitting parameter Dq, and crystal field strength Dq/B. Based on the methodology described by Sing et. all [23], Racah and crystal splitting parameters were calculated using the measured diffuse reflectance spectra and $R_1$-line emission maximum (Table 2). It should be noted that the calculated intersection of $^2E_g$, and $^4T_{2g}$ levels was different for different samples and varied from 1.94(5) to 2.09(5). The calculated values of Dq/B were in the range of 2.14(5) - 2.32(5). The calculated values of Dq/B for all samples were above the intersection, which indicates that the $Cr^{3+}$ ions are located in a strong crystal field. It is of interest that the difference between the calculated crystal field strength Dq/B and the intersection was in close vicinity for all samples and equal to 0.20-0.24 with the exception of Cr30 sample. This indicates that the luminescence spectra should be similar for all samples.

Table 2: Racah parameters, crystal splitting parameter (Dq), crystal field strength (Dq/B), and C/B ratio for $Cr^{3+}$ ions in the samples. Dq/B ($^2E_g$=$^4T_{2g}$) indicates the intersection of $^2E_g$, and $^4T_{2g}$ energy levels.

| denote | Cr, at.% | Cr, $cm^{-3}$ | Dq, $cm^{-1}$ | B, $cm^{-1}$ | C, $cm^{-1}$ | Dq/B | Dq/B ($^2E_g$=$^4T_{2g}$) | C/B |
|---|---|---|---|---|---|---|---|---|
| Cr01 | 0.1 | $2.2 \cdot 10^{19}$ | 1590(5) | 744(15) | 3033(50) | 2.14(5) | 1.94(5) | 4,1(1) |
| Cr05 | 0.5 | $1.1 \cdot 10^{20}$ | 1610(5) | 741(15) | 3002(50) | 2.17(5) | 1.95(5) | 4,1(1) |
| Cr1 | 1 | $2.2 \cdot 10^{20}$ | 1609(5) | 731(15) | 3057(50) | 2.20(5) | 1.97(5) | 4,2(1) |
| Cr3 | 3 | $6.5 \cdot 10^{20}$ | 1604(5) | 713(15) | 3095(50) | 2.25(5) | 2.02(5) | 4,3(1) |
| Cr5 | 5 | $1.1 \cdot 10^{21}$ | 1601(5) | 690(15) | 3142(50) | 2.32(5) | 2.09(5) | 4,6(1) |
| Cr10 | 10 | $2.2 \cdot 10^{21}$ | 1617(5) | 722(15) | 3074(50) | 2.24(5) | 2.00(5) | 4,3(1) |
| Cr30 | 30 | $6.5 \cdot 10^{21}$ | 1629(5) | 702(15) | 3115(50) | 2.32(5) | 2.05(5) | 4,4(1) |

*3.3 Excitation spectra*

$Cr^{3+}$ excitation maps of low-doped samples exhibit a decrease in intensity, while high-doped samples show Gaussian-like behavior with intensity maxima at higher temperatures. Cr:YGG nanocrystals were investigated by collecting $Cr^{3+}$ excitation bands at different temperatures. The $Cr^{3+}$ excitation maps are shown in Fig. 4. Two distinct patterns can be observed: Cr01, Cr05 and Cr1 exhibit a decrease in excitation intensity with increasing temperature, whereas Cr03, Cr5, and Cr10 show an intensity maximum at higher temperatures. The measured temperature dependence of the excitation area for Cr:YGG nanocrystals is presented in Fig. S3. Cr01, Cr05, and Cr1 exhibit a continuous decrease in excitation intensity across the entire temperature range, while Cr03, Cr5, and Cr10 exhibit a gaussian-like dependence, with maxima at ~15 °C, ~50 °C, and ~100 °C, respectively.

The detected excitation patterns correlate with microstructure variations, which can be connected with the redistribution of $Cr^{3+}$ ions between smaller and larger fractions. The collected excitation patterns correlate with the microstructure of Cr:YGG nanocrystals. Cr01, Cr05, and Cr1 exhibit narrower XRD lines than Cr03 and Cr5, which can be attributed in the ratio of smaller to larger fractions. It is possible that the $Cr^{3+}$ concentration not only influences the morphology of the synthesized Cr:YGG nanocrystals but also affects their distribution between the smaller and larger fractions. Interestingly, the excitation map of the Cr10 sample is more similar to those of Cr03, and Cr5 samples, even though its microstructure resembles that of Cr01, Cr05, and Cr1 samples. However, the overall emission intensity of Cr10 is one order of magnitude lower than that of Cr5, indicating strong luminescence quenching.

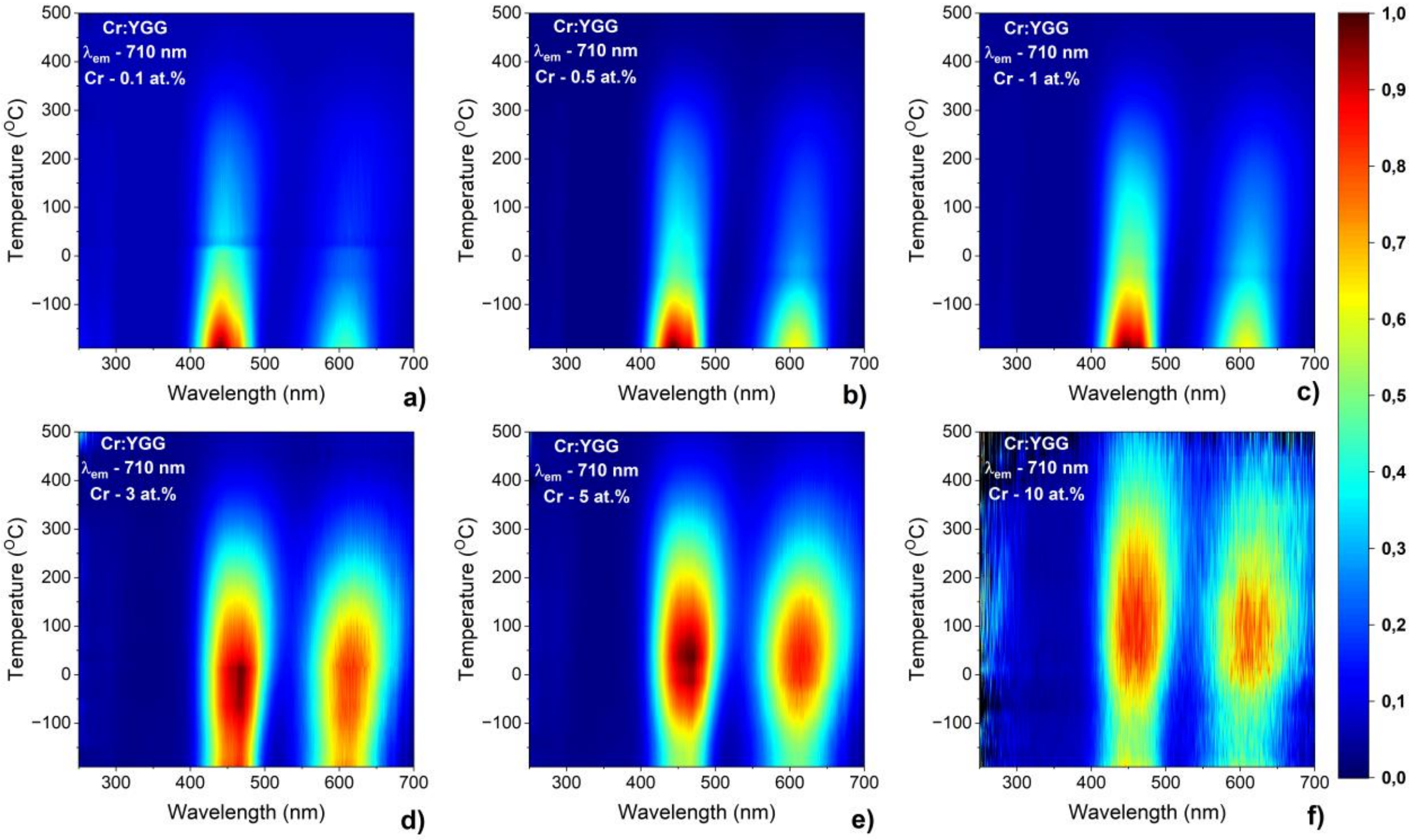

Fig. 4: Temperature dependence of $Cr^{3+}$ excitation bands measured at $\lambda_{em}$ – 710 nm for a) Cr01, b) Cr05, c) Cr1, d) Cr3, e) Cr5, and f) Cr10, respectively.

The recorded excitation spectra exhibit a typical shape for $Cr^{3+}$ ions in garnets; however, the bands are clearly the superposition of several bands. The evolution of the excitation spectra was analyzed by monitoring the excitation band parameters through the deconvolution by a log-normal function [24] using Eq.1:

$$I(\lambda) = H \cdot exp\left\{\frac{-\ln 2}{(\ln p)^2}\left[\ln\left(\frac{\lambda-\lambda_{max}}{\Delta_{FWHM}}\frac{(p^2-1)}{p}+1\right)\right]^2\right\} \quad (1)$$

The log-normal curve (Eq. 1) is an asymmetric Gaussian function, whose shape is defined by four parameters, $\lambda_{max}$, (band maximum) H, (height), $\Delta_{FWHM}$ (half-width), and p (skewness). The recorded excitation spectra exhibit a typical shape for $Cr^{3+}$ ions in garnets, with two main peaks centered at ~16500 $cm^{-1}$ (~600 nm), and 22500 $cm^{-1}$ (~440 nm), corresponding to ${}^4A_{2g}\rightarrow{}^4T_{2g}$, and ${}^4A_{2g}\rightarrow{}^4T_{1g}$, respectively [20]. Additionally, a third weak band appears in the UV region, with an excitation maximum at 35500 $cm^{-1}$ (~280 nm), This band likely corresponds to a charge transfer transition from oxygen to $Cr^{3+}$ ions [25]. Three narrow spin-forbidden transition bands are observed at 14500 $cm^{-1}$ (~690 nm), 14850 $cm^{-1}$ (~670 nm), and 21600 $cm^{-1}$ (~460 nm), corresponding to ${}^4A_{2g}\rightarrow{}^2E_g$, ${}^4A_{2g}\rightarrow{}^2T_{1g}$, and ${}^4A_{2g}\rightarrow{}^2T_{2g}$ transitions, respectively. The shape of the ${}^4A_{2g}\rightarrow{}^4T_{1g}$ excitation band suggests the presence of at least two separate sub-bands centered at 21500 $cm^{-1}$ (~470 nm), 22500 $cm^{-1}$ (~440 nm). It is also likely that the ${}^4A_{2g}\rightarrow{}^4T_{2g}$ band consists of two overlapping sub-bands, although they cannot be clearly distinguished due to strong overlap.

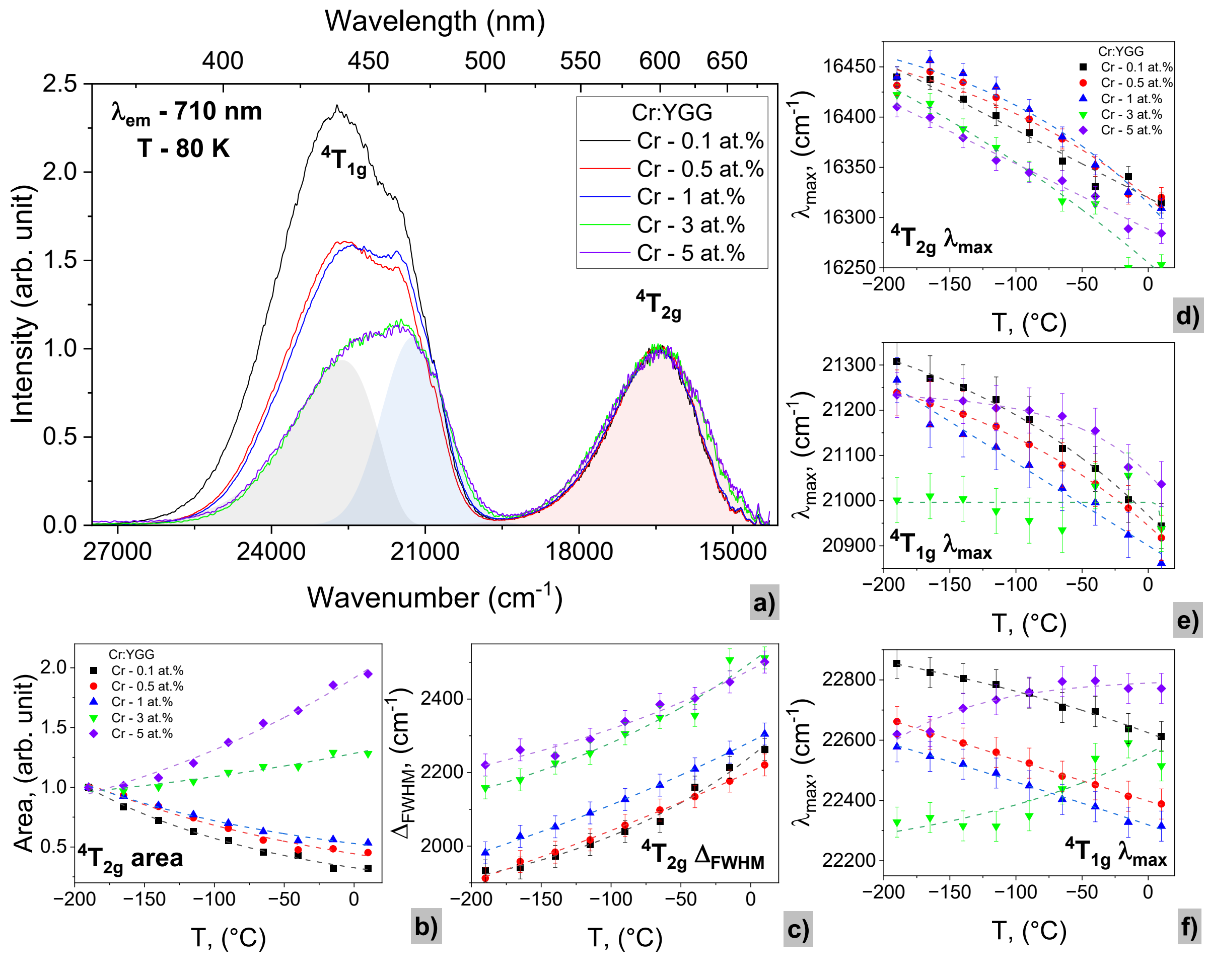

Fig. 5: a) Cr:YGG excitation spectra measured at $\lambda_{em}$ – 710 nm and T – 80K. Temperature dependence of $Cr^{3+}$ excitation band parameters: b) ${}^4T_{2g}$ band area; c) $\Delta_{FWHM}$ of the ${}^4T_{2g}$ band; d) $\lambda_{max}$ of the ${}^4T_{2g}$ band; e) $\lambda_{max}$ of the first ${}^4T_{1g}$ band; f) $\lambda_{max}$ of the second ${}^4T_{1g}$ band.

The different patterns for the temperature dependence of excitation band intensity were found for the low and high-doped samples. The temperature evolution of the excitation band parameters for Cr:YGG nanocrystalline is shown in Fig. 5. Two distinct trends can be identified: the first group includes Cr01, Cr05, and Cr1 samples, while the second group consists of Cr3 and Cr5 samples. Increasing the temperature from -190 °C to 10 °C leads to a decrease in the intensity of the $^4T_{2g}$ excitation band for Cr01, Cr05, and Cr1, whereas an increase in intensity is observed for Cr3 and Cr5 (Fig. 5(b)). The overall evolution of the $Cr^{3+}$ excitation band intensity is presented in Fig. S3. The $\Delta_{FWHM}$ of the $^4T_{2g}$ excitation band increases with temperature at a similar rate for all samples (Fig. 5(c)), but the initial values differ. At 80 K, the measured $\Delta_{FWHM}$ of the $^4A_{2g}\rightarrow{}^4T_{2g}$ excitation band was 1932(30), 1912(30), 1981(30), 2158(30), 2220(30), for Cr01. Cr05, Cr1, Cr3, and Cr5, respectively. The measured $\Delta_{FWHM}$ of the excitation bands correlates with the $\Delta_{FWHM}$ of XRD peaks, indicating a less uniform distribution of the local crystal field around $Cr^{3+}$ ions in Cr3, and Cr5 samples.

Different patterns were detected for the evolution of the band excitation maximums for high and low doped samples, and might be caused by the change of the chromium redistribution between the different garnet fractions with the increase of $Cr^{3+}$ concentration. Increasing the temperature resulted in a decrease of the excitation maximum of the $^4T_{2g}$ band for all samples. However, Cr3, and Cr5 samples exhibited a ~50 $cm^{-1}$ redshift compared to Cr01. Cr05, and Cr1 samples (Fig. 5(d)). At 80 K, the excitation maximum of the $^4T_{2g}$ band was 16314(10), 16320(10), 16309(10), 16253(10), and 16284(10) for Cr01. Cr05, Cr1, Cr3, and Cr5, respectively. Due to the overlapping of the two $^2T_{1g}$ excitation sub-bands, it is difficult to precisely determine their excitation peaks parameters. This issue became particularly challenging at high temperatures due to band broadening. The two bands, centered at 21500 $cm^{-1}$ and 22500 $cm^{-1}$, are labeled as $^4T_{1g}(1)$, and $^4T_{1g}(2)$, respectively. The temperature dependence of these bands is shown in Fig. 5(e,f) and Fig. S4. These bands likely originate from $Cr^{3+}$ ions in different garnet fractions (see microstructure section for details). As the temperature increased, the excitation maximum of the $^4T_{1g}(1)$ band decreased, but no clear trend was observed between Cr01. Cr05, Cr1, and Cr3, Cr5 samples. For the $^4T_{1g}(2)$ band, increasing the sample temperature resulted in a decrease in the band maximum for Cr01. Cr05, Cr1. The opposite trend was observed for Cr3, and Cr5, where the excitation maximum increased with temperature.

It is expected to be detected a redshift of the excitation bands with the increase in temperature, which was the case for the low-concentrated samples. A large Stokes shift and broad optical bands indicate a significant difference in ion–lattice coupling for both $^4T_{2g}$ and $^4T_{1g}$ excited states relative to the ground state $^4A_{2g}$ [7]. This suggests that the $^4A_{2g}\rightarrow^2T_{1g}$ and $^4A_{2g}\rightarrow^2T_{2g}$ electronic transitions are phonon-assisted, meaning that the electron transition ends in a higher vibronic state. As temperature increases, the population of vibronic states also increases, requiring less energy to populate the $^2T_{1g}$ and $^2T_{2g}$ states. This causes a redshift in the excitation bands. A redshift in the $^4T_{2g}$ and $^4T_{1g}$ bands was observed in the samples, except for the $^4A_{2g}\rightarrow^2T_{1g}$ transition in Cr3 and Cr5.

For a single $Cr^{3+}$ ion, an increase in the temperature should lead to a decrease in the excitation band area across all measured temperatures, which wasn't the case for highly doped samples. Electron transitions from the emission level occur through both radiative and nonradiative processes. If only the radiative recombination path were present, the area under the excitation bands would remain constant, with only a shift in $\lambda_{max}$ and $\Delta_{FWHM}$ being observed [7]. However, as the temperature increases, nonradiative recombination pathways become more significant. This led to a reduction in the excitation band area across all temperatures. This expected behavior was observed for the $^4A_{2g}\rightarrow^2T_{1g}$ and $^4A_{2g}\rightarrow^2T_{2g}$ excitation bands in Cr01, Cr05, and Cr1. In constant, Cr3 and Cr5 exhibit the opposite trend in the temperature range from -80 °C to 0 °C, suggesting a different temperature-dependent recombination mechanism in these samples.

The detected deviation from the expected behavior of excitation band parameters with the temperature might be due to enhanced energy transfer between $Cr^{3+}$ ions in different local environments. A possible explanation for the increase in excitation bands could be related to the enhanced efficiency of energy transfer between $Cr^{3+}$ ions [26]. One characteristic feature of $Cr^{3+}$ ions is their ability to efficiently transfer energy both to other ions and among themselves. The presence of two distinct fractions of nanocrystals (Fig. 1(b)) suggests the existence of various $(CrO_6)^{9-}$ optically active centers. The observed increase in excitation band intensity with temperature is likely due to the redistribution of the absorbed energy among $Cr^{3+}$ ions located in different $CrO_6)^{9-}$ centers.

*3.4 Emission spectra*

Despite the similarity in the Racah and crystal field parameters, the luminescence spectra of the samples varied with concentration, suggesting the presence of multiple $Cr^{3+}$ optical centers. The low-temperature $Cr^{3+}$ emission spectra of Cr:YGG nanocrystals consist of the spin-forbidden

$^2E_g \rightarrow ^4A_{2g}$ transition with vibronic sidebands and the spin-allowed broadband $^4T_{2g} \rightarrow ^4A_{2g}$ transition (Fig. 6) [27,28]. The variation in $Cr^{3+}$ concentration has little effect on the position of R-lines and vibronic sidebands of the $^2E_g \rightarrow ^4A_{2g}$ transition [25,29]. However, an increase in $Cr^{3+}$ concentration causes a redshift of the emission maximum of the $^4T_{2g} \rightarrow ^4A_{2g}$ broadband transition. The collected emission spectra suggest the presence of at least two distinct types of $(CrO_6)^{9-}$ optically active centers. For example, the emission spectra of Cr05 and Cr10 samples exhibit a broadband emission line with a maximum exceeding 800 nm, which originates from $Cr^{3+}$ ions in a weak crystal field ($Cr^{3+}_{WEAK}$) [14]. However, $Cr^{3+}_{WEAK}$ ions are typically characterized by the absence of R-liens emission, which is not the case in our samples (Fig. 6). This suggests that the Cr03,Cr05, and Cr1 samples contain both $Cr^{3+}_{WEAK}$ and $Cr^{3+}_{STRONG}$ ions ($Cr^{3+}_{STRONG}$ is $Cr^{3+}$ in a strong crystal filed).

The samples with the lower concentration of $Cr^{3+}$ ions exhibit a similar luminescence spectrum to bulk Cr:YGG, while the highly doped samples show a broad luminescence band in NIR region typical for $Cr^{3+}$ ions in a weak crystal field. With the exception of vibronic sidebands, $Cr^{3+}$ emission spectra can be fitted using three log-normal functions, representing $^4T_{2g} \rightarrow ^4A_{2g}$ broadband transition and two narrow R-lines. Notably, the calculated Dq/B for all samples is ~0.2 higher than the intersection point (Table 2). This suggests that the lowest excited energy level of $Cr^{3+}$ ions is $^2E_g$. By analyzing the changes in line parameters, the samples investigated can be divided into two distinct groups:

- Group 1: Cr0.1, Cr0.5, and Cr1, which exhibit emission spectra similar to bulk Cr:YGG.
- Group 2: Cr3, Cr5, and Cr10, which are characterized by the presence of the $^4T_{2g} \rightarrow ^4A_{2g}$ emission band in the infrared region.

High temperature luminescence spectra of the samples exhibited similar shapes and positions, indicating a similar environment around the emitting centers. The observed features of the emission spectra correlate with the measured temperature dependence of the excitation bands (Fig. 4). Notably, the luminescence spectra of $Cr^{3+}$:YGG nanopowders at 100 °C exhibit a similar shape (Fig. 6(b)). These spectra show no R-lines emission and display only $^4T_{2g} \rightarrow ^4A_{2g}$ broadband emission, with $\lambda_{max}$ ~725 nm. This suggests that as the sample temperature increases, the emission from $Cr^{3+}_{WEAK}$ disappears.

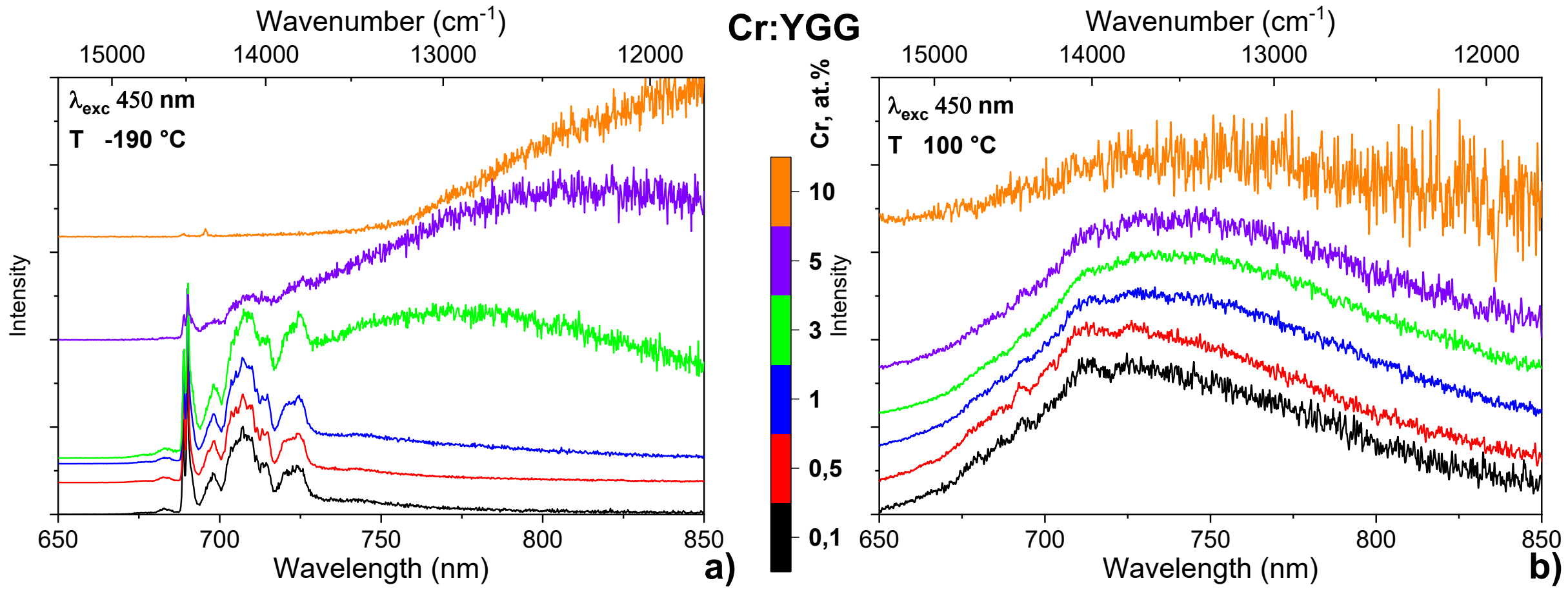

Fig. 6: Emission spectra of $Cr^{3+}$ ions in Cr:YGG nanocrystals measured at $\lambda_{exc}$ – 450 nm and a) T – -190 °C, T – 100 °C.

The temperature dependence of $Cr^{3+}$ luminescence behavior is determined, among other factors, by the thermal coupling between the two lowest energy states. Irradiation of $Cr^{3+}$:YGG nanocrystals with 450 nm light induces the $^4A_{2g}\rightarrow{}^4T_{1g}$ electronic transition, followed by nonradiative relaxation to the $^2E_g$ state. Two possible recombination pathways exist:

1. Radiative recombination to the $^4A_{2g}$ ground state, producing characteristic R-lines emission.
2. Thermalization of $^4T_{2g}$ state with the following radiative or nonradiative recombination.

The increase in temperature caused the redistribution of the absorbed energy among the lowest emission levels, changing their contribution to luminescence spectra. The rate of the $^2E_g\rightarrow{}^4T_{2g}$ transition depends on the energy difference between these states and the temperature [7]. As the temperature increases, the $^2E_g\rightarrow{}^4T_{2g}$ transition rate rises, reducing the population of the $^2E_g$ state and enhancing emission from the $^4T_{2g}$ state [30]. This trend was observed in Cr0.1, Cr0.5, Cr1 samples. Specifically:

- The increase in temperature resulted in a decrease in $^2E_g\rightarrow{}^4A_{2g}$ emission intensity (Fig. 8(d)), which dropped to zero at 100 °C.
- The increase in the $^4T_{2g}\rightarrow{}^4A_{2g}$ emission intensity (Fig. 8(a)) due to enhanced energy transfer from $^2E_g$ to $^4T_{2g}$ energy levels.

The decrease in the R-lines emission and U-shape of broadband emission intensity with temperature is characteristic of a single type of $(CrO_6)^{9-}$ optically active center.

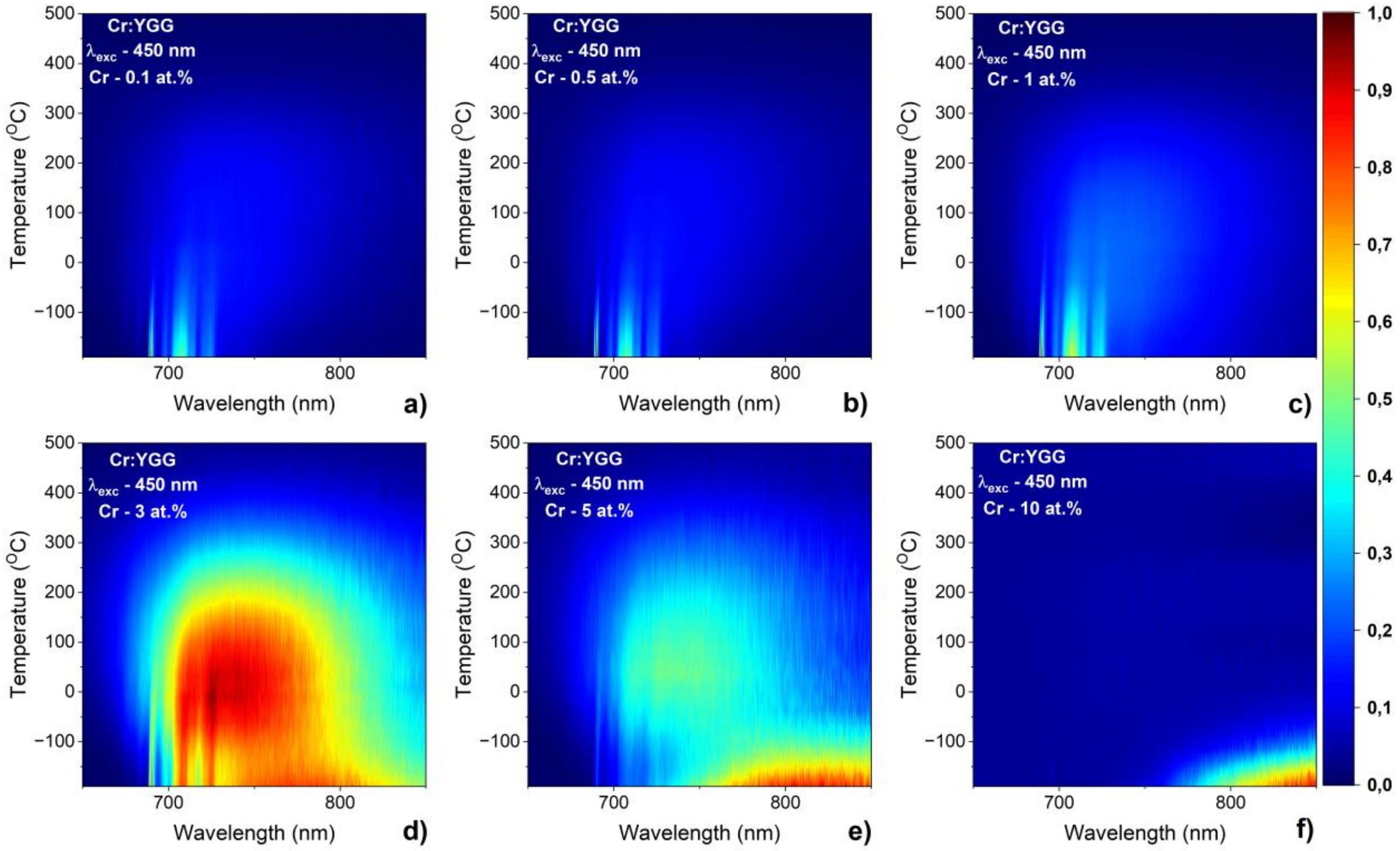

Fig. 7: Temperature dependence of $Cr^{3+}$ emission for a) Cr0.1, b) Cr0.5, c) Cr1, d) Cr3, e) Cr5, and f) Cr10 measured at $\lambda_{exc}$ – 450 nm.

The change in luminescence band parameters for high-doped samples suggests the presence of multiple optically active centers. A different trend was observed for the Cr3, Cr5, and Cr10 samples.

- Cr3: The $^4T_{2g}\rightarrow{}^4A_{2g}$ emission intensity remained constant in the temperature range of -190 °C to 50 °C, followed by gradual decrease (Fig. 8(a)).
- Cr5: The $^4T_{2g}\rightarrow{}^4A_{2g}$ emission intensity decreased from -190 °C to -50 °C, remained unchanged up to 100 °C, and then declined (Fig. 8(a)).

The detected pattern in the change of luminescence parameters for high-doped samples is different from the one expected for a single type of $Cr^{3+}$ ions. For $Cr^{3+}$ ions in a strong crystal field (such as in Cr3, and Cr5 samples), a U-shaped dependence of $^4T_{2g}\rightarrow{}^4A_{2g}$ emission intensity is expected. However, an unusual pattern was observed in the temperature-dependent behavior of the $^2E_g\rightarrow{}^4A_{2g}$ emission intensity for Cr3 and Cr5.

- Cr3: The $^2E_g\rightarrow{}^4A_{2g}$ intensity decreased from -190 °C to -60 °C, remained constant from -60 °C to -10 °C, and then declined further (Fig. 8(d)).

- Cr5: The $^2E_g \rightarrow ^4A_{2g}$ intensity increased from -190 °C to -100 °C, followed by a gradual decrease (Fig. 8(d)).

The detected patterns suggest the presence of several types of $Cr^{3+}$ optically active centers. For a single type of $(CrO_6)^{9-}$, the R-lines ($^2E_g \rightarrow ^4A_{2g}$) emission intensity should change in only one direction with increasing temperature: *unchanged → decrease → absence*. The observed trends *decrease → unchanged → decrease* (Cr3) and *increase → decrease* (Cr5) are inconsistent with the expected behavior, suggesting the presence of multiple$(CrO_6)^{9-}$ optically active centers.

It is possible that two types of centers are involved in Cr:YGG samples luminescence, $Cr^{3+}$ ions in a strong crystal field and $Cr^{3+}$ ions in a weak crystal field. The observed changes in both R-lines and broadband emission intensity can be explained by the presence of two types of $(CrO_6)^{9-}$ optically active centers: $Cr^{3+}_{STRONG}$, and $Cr^{3+}_{WEAK}$. As the temperature increases, the $^4T_{2g} \rightarrow ^4A_{2g}$ emission intensity is expected to:

- Decrease for $Cr^{3+}_{WEAK}$ ions.
- Exhibits a reverse U-shape trend for $Cr^{3+}_{STRONG}$ ions.

The detected patterns in the change of luminescence and excitation spectra could be explained by the change in contribution to luminescence due to the energy transfer process between these centers. The measured excitation and luminescence spectra suggest that both $Cr^{3+}_{WEAK}$ and $Cr^{3+}_{STRONG}$ ions were present in Cr3, and Cr5. The overlapping emission patterns from these two types of centers likely explain the observed changes in the $^4T_{2g} \rightarrow ^4A_{2g}$ emission intensity. However, this explanation does not fully account for the changes in R-lines emission intensity. The increase in R-lines emission intensity is most likely due to the energy transfer between multiple $(CrO_6)^{9-}$ optically active centers [10].

The recorded decrease in the width of luminescence bands at some temperature interval for the high-doped samples is in favor of the proposed earlier multicenter nature of the Cr:YGG luminescence. The observed trend in the change of FWHM with temperature for the $^4T_{2g} \rightarrow ^4A_{2g}$ emission also suggests the presence of multiple emission centers in Cr3, and Cr5 (Fig. 8(b)). The variation in the bandwidth of the absorption and emission bands with temperature follows the equation [7]:

$$\Delta E(T) \approx \Delta E(0) \sqrt{coth\left(\frac{\mathrm{h}\Omega}{2kT}\right)}$$

where ΔE(0) is the bandwidth at 0 K, and hΩ is the energy of the coupling phonon. This equation predicts an increase in the FWHM with temperature, which is consistent with the behavior observed in Cr0.1, Cr0.5, and Cr1. For $Cr^{3+}$:GGG nanoceramics, an increase in temperature (10 °C→ 500 °C) results in a linear increase in the FWHM for the $^4T_{2g}$→$^4A_{2g}$ emission [10]. The other pattern was found for Cr3 and Cr5:

- Cr3: A decrease in FWHM for the $^4T_{2g}$→$^4A_{2g}$ emission (-190 °C → 0 °C), followed by a match with the common trend (Fig. 8(b)).
- Cr5: An increase in FWHM for the $^4T_{2g}$→$^4A_{2g}$ emission (-190 °C → -110 °C), followed by a decrease (-115 °C → 20 °C), and then a match with the common trend (Fig. 8(b)).

Variation in the temperature change contributes to the luminescence of high-doped samples from different $Cr^{3+}$ ions, which explains the variation in the width of the luminescence. This behavior of the Cr3 and Cr5 is typical of multiple $(CrO_6)^{9-}$ centers. The luminescence spectra of Cr3, and Cr5 are dominated by the emission from $Cr^{3+}_{WEAK}$ ions (-190 °C → 20 °C), white at higher temperatures, the emission from $Cr^{3+}_{STRONG}$ ions become the dominant feature. On the other hand, the Cr0.1, Cr0.5, and Cr1 samples are characterized by emission solely from $Cr^{3+}_{STRONG}$ ions across the entire temperature range.

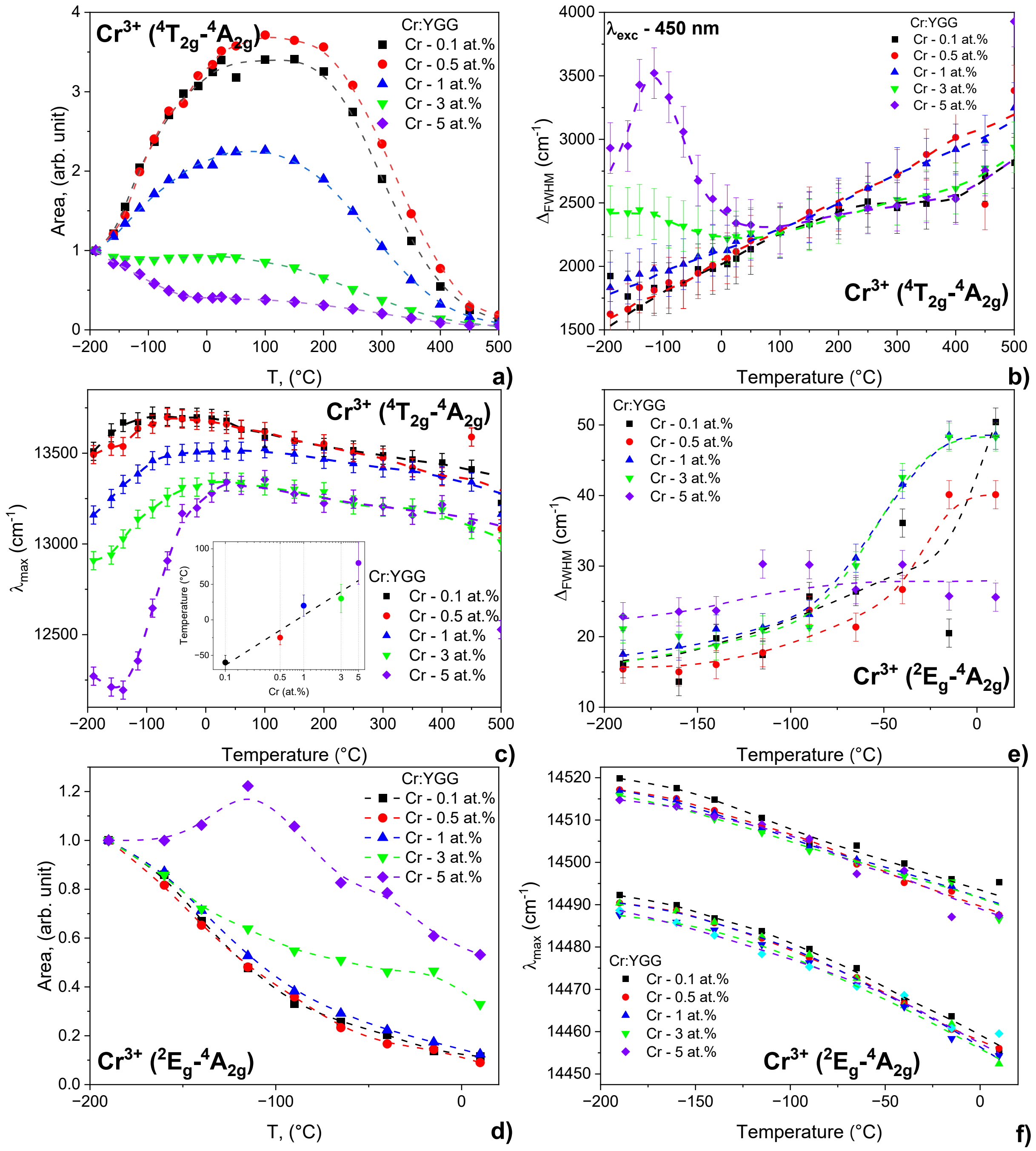

Fig. 8: Temperature dependence of $Cr^{3+}$ emission band parameters: $^4T_{2g}\rightarrow{}^4A_{2g}$ a) area; b) $\Delta_{FWHM}$; c) $\lambda_{max}$; and $^2E_g\rightarrow{}^4A_{2g}$ d) R-lines area, d) $R_1$-line $\Delta_{FWHM}$; e) $\lambda_{max}$.

The detected variation of the broadband emission maximum with temperature suggests the contribution from the $Cr^{3+}$ ions in both weak and strong crystal fields to luminescence for all samples. The increase in temperature caused a shift of $\lambda_{max}$ for the $^4T_{2g}\rightarrow{}^4A_{2g}$ broadband emission (Fig. 8(c)). This trend was observed across all samples, with an initial blueshift of $\lambda_{max}$ for $T < T_C$, followed by a redshift of $\lambda_{max}$ for $T > T_C$. In the case of $Cr^{3+}$:GGG nanoceramics, the increase in temperature (10 °C→ 500 °C) resulted in a linear decrease in the $\lambda_{max}$ [10]. The observed rise in $\lambda_{max}$ may be attributed to the presence of the $Cr^{3+}_{WEAK}$ component in the emission spectra. Interestingly, the $T_C$ for Cr3 and Cr5 (Fig. 8(c)) coincides closely with the temperatures at which the FWHM match the common trend (Fig. 8(b)). This supports the conclusion that the $Cr^{3+}$ luminescence consists of a superposition of $Cr^{3+}_{WEAK}$, and $Cr^{3+}_{STRONG}$, each contributing at different percentage. The difference in $\lambda_{max}$ for the $^4T_{2g}$→$^4A_{2g}$ emission at -190 °C indicates a decrease in the local Dq/B around $Cr^{3+}_{WEAK}$ ions as the $Cr^{3+}$ concentration increases. The $\lambda_{max}$ values for the $^4T_{2g}$→$^4A_{2g}$ emission at -190 °C were ~13500 $cm^{-1}$ (Cr0.1) ~13500 $cm^{-1}$ (Cr0.5), ~13150 $cm^{-1}$ (Cr1), ~12900 $cm^{-1}$ (Cr3), and ~12200 $cm^{-1}$ (Cr5).

Predicting the spectral position of broadband emission of $Cr^{3+}$ ions is challenging; therefore, the detected pattern in the luminescence behavior of the low-doped sample can be attributed solely to a single type of $Cr^{3+}$ ions. Based on the electronic structure of $Cr^{3+}$ ions, predicting the spectral position of the $^4T_{2g}$↔$^4A_{2g}$ transition in $Cr^{3+}$-doped garnets at different temperatures is challenging. This is due to the difference in the ion–lattice coupling between the $^4T_{2g}$ excited state and the $^4A_{2g}$ ground state. This makes the electron transitions between these levels strongly dependent on the configurational coordinate Q. In other words, the $^4A_{2g}$→$^4T_{2g}$ electron transitions start from the minimum of the $^4A_{2g}$ parabola and end at a higher vibronic state of the $^4T_{2g}$ levels. Consequently, the $^4T_{2g}$→$^4A_{2g}$ transition begins from the minimum of the $^4T_{2g}$ parabola and ends at a higher vibronic state of the $^4A_{2g}$ energy levels. As the temperature increases, the population of higher vibronic states increases, causing the electron transitions to begin or end in higher vibronic states. The result is a shift in the spectral positions of the respective bands. Therefore, the observed change in the $\lambda_{max}$ for the $^4T_{2g}$→$^4A_{2g}$ in Cr0.1, Cr0.5, and Cr1 can be attributed solely to the $Cr^{3+}_{STRONG}$ ions, without the participation of $Cr^{3+}_{WEAK}$ ions in the luminescence.

The change in the parameters of R-lines suggests a similar local crystal field environment around the “strong crystal field” part of $Cr^{3+}$ centers. $Cr^{3+}$ concentration has a minimal effect on R-lines emission parameters, with temperature broadening and a redshift in the R1-line, indicating similar Dq/B for strong crystal field $Cr^{3+}$ ions. With the exception of emission intensity, variations in $Cr^{3+}$ concentration have minimal impact on the R-lines emission parameters. An increase in temperature leads to broadening of the $R_1$-line emission, with FWHM changes from 15 $cm^{-1}$ to 45 $cm^{-1}$ (Fig. 8(e)). No pattern of the FWHM changes was found for $R_1$-line (except for Cr5). R-lines position exhibits a redshift of ~30 $cm^{-1}$ as the temperature increases from -190 °C to 0 °C (Fig. 8(f)), following the same trend for all samples.

The variation in $\lambda_{max}$ remained within 5 $cm^{-1}$, indicating that the $Cr^{3+}_{STRONG}$ ions exhibit a similar Dq/B across the samples.

The recorded overall $Cr^{3+}$ emission exhibits an increase at some temperature interval, which suggests the presence of both types of $Cr^{3+}$ ions in all samples. Fig. 9 shows the variation in overall $Cr^{3+}$ emission intensity with temperature.

- Cr0.1, Cr0.5, and Cr1 exhibit a 10% *increase* in emission intensity (-190 °C → 100 °C), followed by a decline.
- Cr3 and Cr5 display a *decline* → *increase* → *decline* pattern of $Cr^{3+}$ luminescence intensity.
- Cr10 shows a steep *decline* in luminescence intensity from 190 °C to -50 °C.

In an ideal system, temperature increase leads to energy redistribution between $^2E_g$ and $^4T_{2g}$ states, while the total emission intensity should remain unchanged [7]. However, energy transfer processes and nonradiative relaxation processes introduce additional loss channels, reducing overall $Cr^{3+}$ emission. The nonradiative relaxation rate increases with temperature. Therefore, for single $Cr^{3+}$ ions, it is expected emission trend follows a single pattern: *unchanged* → *decrease*.

- Cr0.1, Cr0.5, Cr1, and Cr3 exhibit an increase in emission intensity (-190 °C → 0 °C) (Fig. 9).
- Cr5 and Cr10 follow a *decrease* → *unchanged* emission trend (-190 °C → 0 °C) (Fig. 9).

These variations in emission intensity suggest the presence of multiple $(CrO_6)^{9-}$ optically active centers, even in Cr0.1, Cr0.5, Cr1 samples.

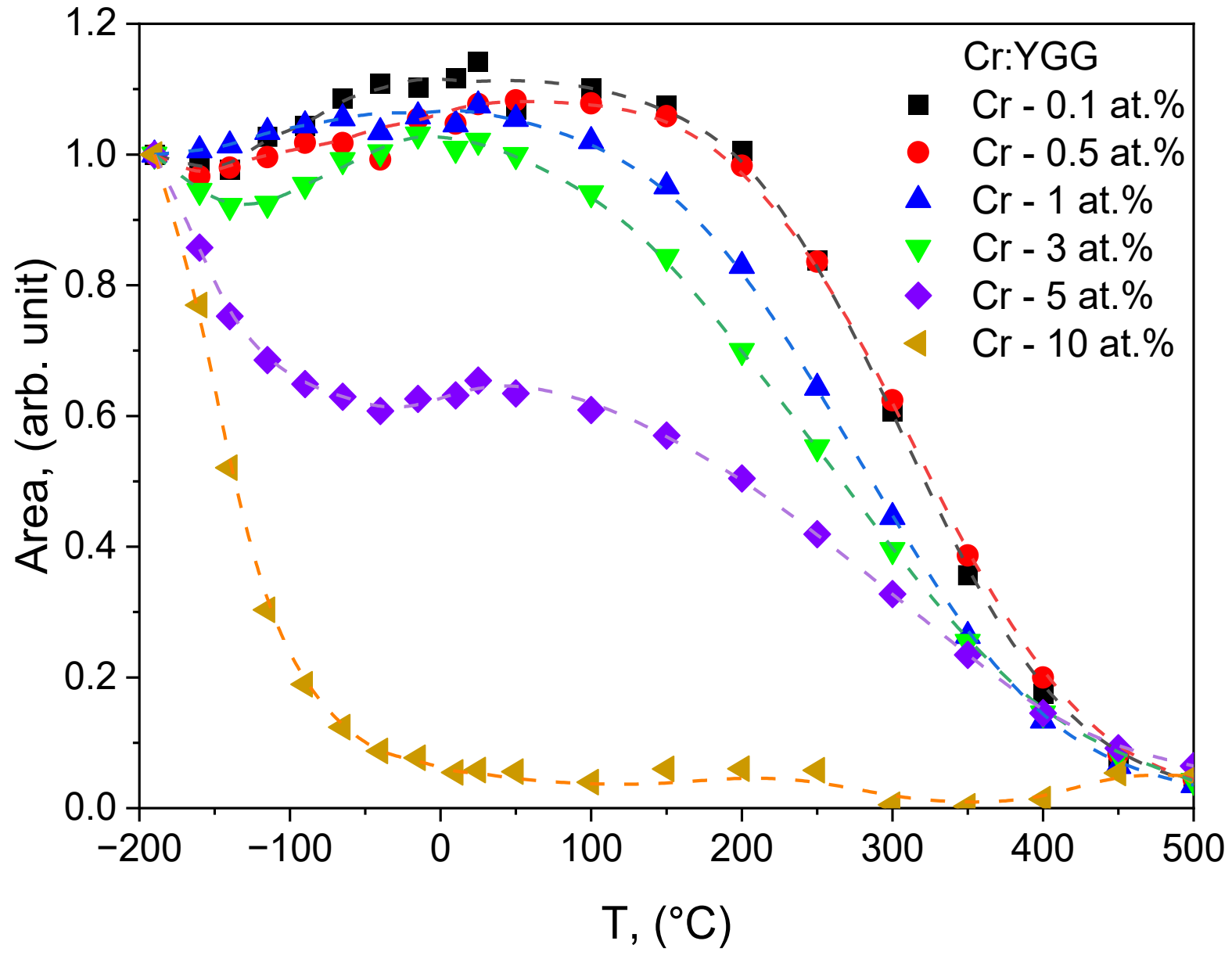

Fig. 9: Temperature dependence of $Cr^{3+}$ emission area for Cr:YGG nanopowders. The collected values were normalized on the $Cr^{3+}$ emission at -190 °C.

The recorded variation in the overall $Cr^{3+}$ luminescence for low-concentrated samples might also be explained by the change in absorption with the change in temperature. The increase in $Cr^{3+}$ emission intensity was previously observed in $Cr^{3+}$:YAB and $Cr^{3+}$:GAB single crystals [1]. In those studies, an increase in $Cr^{3+}$ luminescence intensity was recorded at T > 0 °C and was attributed to an increase in radiative transition rates from the $^4T_{2g}$ energy level, combined with a temperature-independent nonradiative transition rate from the same level. Alternatively, it was proposed that changes in of $^4A_{2g}\rightarrow^4T_{1g}$ absorption could also contribute to the increase in emission intensity [2]. These explanations are applicable to the Cr0.1, Cr0.5, Cr1, and Cr3 samples, but do not account for the observed changes in $Cr^{3+}$ emission in Cr5, and Cr10 samples. The measured emission intensity trends in Cr5, and Cr10 align well with the proposed model of multiple $(CrO_6)^{9-}$ optically active centers. Furthermore, the detected variation in overall $Cr^{3+}$ emission in Cr0.1, Cr0.5, Cr1, and Cr3 also suggests the presence of multiple $(CrO_6)^{9-}$ center.

$Cr^{3+}$ emission changes suggest the presence of multiple $(CrO_6)^{9-}$ centers; however, the excitation spectrum suggests a small concentration of $Cr^{3+}$ ions in a low crystal field, despite their dominant contribution to luminescence in some cases. While the changes in $Cr^{3+}$ ions emission indicate the presence of multiple $(CrO_6)^{9-}$ optical active centers, they do not provide direct information about their concentrations. For instance, the low temperature luminescence spectra of Cr10 are dominated by broadband emission with $\lambda_{max}$ – 850 nm, which originates from $Cr^{3+}_{WEAK}$ ions. However, this does not necessary imply that $Cr^{3+}_{WEAK}$ ions are the majority fraction of $Cr^{3+}$ ions. It is possible that the concentration of $Cr^{3+}_{WEAK}$ is relatively low. Their emission is detected due to the energy transfer from $Cr^{3+}_{STRONG}$ to $Cr^{3+}_{WEAK}$, similar to the behavior observed in $Cr^{3+}$:GGG nanoceramics [10].

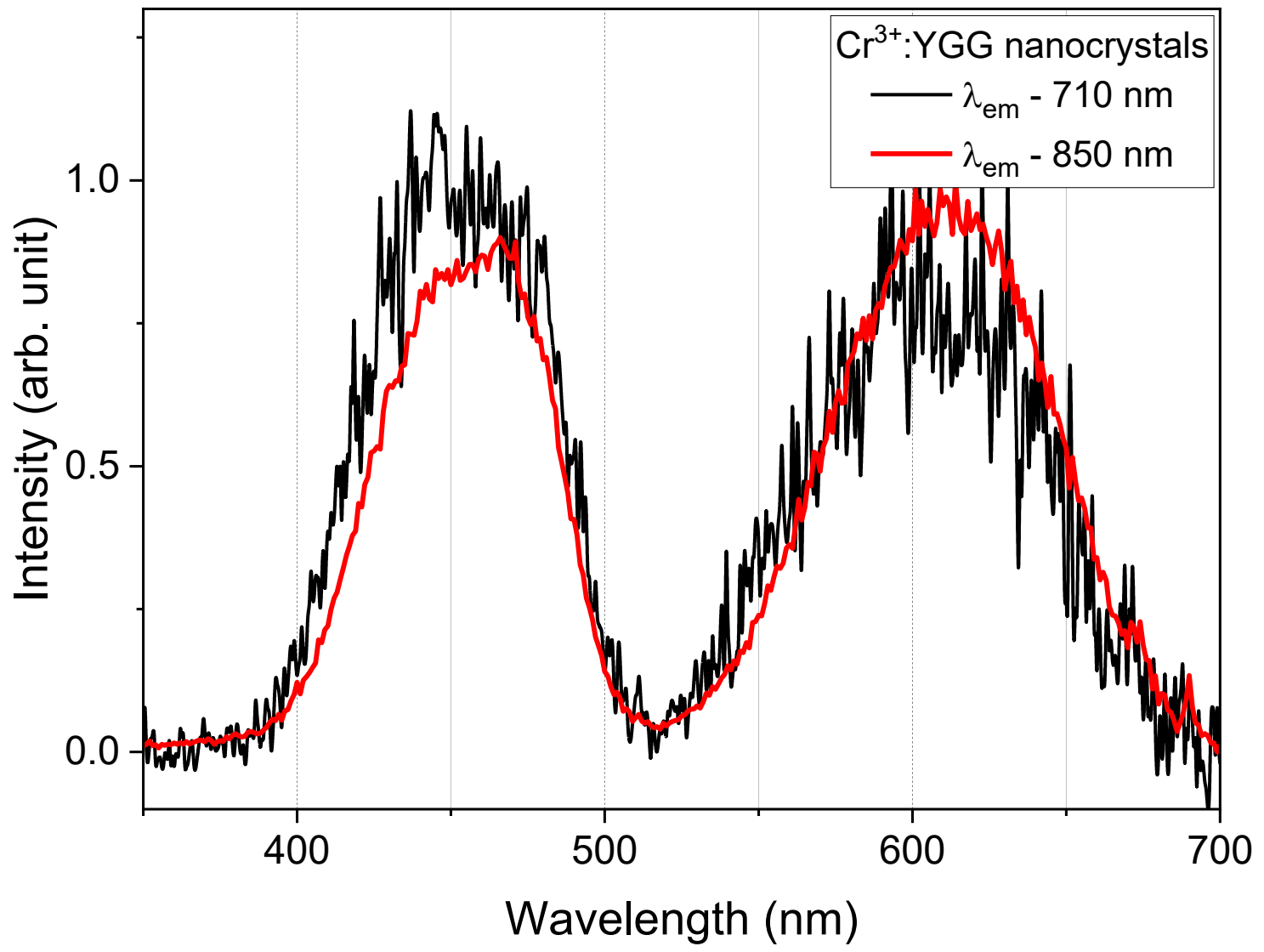

Fig 10: Excitation spectra of $Cr^{3+}$ ions in Cr:YGG nanocrystals (Cr10) measured at -190 °C and $\lambda_{em}$ – 710 nm (black line), and $\lambda_{em}$ – 850 nm (red line).

The presence of broadband emission with emission maximum larger than 675 nm, as in the case of high-doped samples, clearly confirms the presence of $Cr^{3+}$ ions in a low crystal field. Changes in the crystal field strength around $Cr^{3+}$ ions lead to variations in both excitation and emission bands. For example, $Cr^{3+}$-doped single crystals exhibit different Dq/B values and spectral positions for their transitions [2].

- $Cr^{3+}$:LuSGG (Dq/B – 2.57) → $\lambda_{max}$ ($^4A_{2g}$→$^4T_{2g}$) - 607 nm, $\lambda_{max}$ ($^4T_{2g}$→$^4A_{2g}$) at 720 nm
- $Cr^{3+}$:YSGG (Dq/B – 2.50), → $\lambda_{max}$ ($^4A_{2g}$→$^4T_{2g}$) - 618 nm, $\lambda_{max}$ ($^4T_{2g}$→$^4A_{2g}$) at 728 nm
- $Cr^{3+}$:GSGG (Dq/B – 2.45), → $\lambda_{max}$ ($^4A_{2g}$→$^4T_{2g}$) - 636 nm, $\lambda_{max}$ ($^4T_{2g}$→$^4A_{2g}$) at 748 nm
- $Cr^{3+}$:LaSGG (Dq/B – 2.27) → $\lambda_{max}$ ($^4A_{2g}$→$^4T_{2g}$) - 675 nm, $\lambda_{max}$ ($^4T_{2g}$→$^4A_{2g}$) at 805 nm

Based on these trends, it can be assumed that the $(CrO_6)^{9-}$ optical active center in YGG nanocrystals with $^4T_{2g}$ emission at $\lambda_{max}$ – 850 should exhibit an excitation band corresponding to the $^4A_{2g}$→$^4T_{2g}$ transition at $\lambda_{max}$ > 675 nm. Our earliest studies confirmed that $Cr^{3+}$ ions with minor differences in the local crystal field within the same sample (e.g., Cr:YAG transparent ceramic) exhibit a shift in excitation bands [31].

The similarity in the excitation spectra at low temperatures for luminescence from different types of $Cr^{3+}$ ions suggests that absorption occurs at ions in a strong crystal field. Fig. 10 presents the excitation spectra of the Cr10 at -190 °C, recorded for $\lambda_{em}$ – 850 nm, and $\lambda_{em}$ – 710 nm, corresponding to the emission of $Cr^{3+}_{WEAK}$ and $Cr^{3+}_{STRONG}$ ions, respectively. The spectra reveal a similar spectral shape, with the $^4T_{2g}$ excitation band located at 606 nm ($\lambda_{em}$ – 710 nm) and 612 nm ($\lambda_{em}$ – 850 nm). However, $Cr^{3+}$ ions, responsible for emission at $\lambda_{max}$ - 850 nm, should not possess a $^4T_{2g}$ excitation band at 612 nm. This suggests that both excitation spectra correspond to the absorption of $Cr^{3+}_{STRONG}$ ions.

The population of low crystal field $Cr^{3+}$ ions emission level occurs due to the energy transfer from strong crystal field $Cr^{3+}$ ions. The measured emission spectra of Cr10 at -190 °C (Fig. 6(a)) confirm the presence of at least two types of $(CrO_6)^{9-}$ optically active centers:

- Strong crystal field sites, indicated by the presence of $^2E_g \rightarrow {^4A_{2g}}$ emission at $\lambda_{em}$ – 688 nm ($R_2$-line) and 690 nm ($R_1$-line).
- Weak crystal field sites, characterized by $^4T_{2g} \rightarrow {^4A_{2g}}$ emission at $\lambda_{em}$ – 850 nm.

In contrast, the excitation spectra reveal only $Cr^{3+}_{STRONG}$ ions (Fig. 10), indicating that the 850 nm-centered emission band in Cr10 occurs due to energy transfer from $Cr^{3+}_{STRONG}$ ions to $Cr^{3+}_{WEAK}$ ions. Furthermore, the absence of an additional excitation band at $\lambda_{max}$ > 675 nm, which would be expected for $Cr^{3+}$ ions with $\lambda_{em}$ – 850 nm [32], suggest that the concentration of $Cr^{3+}_{WEAK}$ ions is several orders of magnitude lover than that of $Cr^{3+}_{STRONG}$ ions.

*3.5 Luminescence decay*

Fluorescence decay curves of Cr:YGG nanocrystals show two exponential decay components, with the fast decay possibly from $Cr^{3+}$ ions in dodecahedral sites, though the cause remains unclear. The influence of temperature on the lifetime of $Cr^{3+}$ ions was investigated. Fig. S5 presents the measured fluorescence decay curves of Cr:YGG nanocrystals at $\lambda_{exc}$ – 450 nm and $\lambda_{em}$ – 690 nm. The decay curves of $Cr^{3+}$ ions are typically fitted by two exponential functions [30,33]:

$$I = I_0 + A_1 e^{(-t/\tau_1)} + A_2 e^{(-t/\tau_2)} \qquad (3)$$

where I is the luminescence intensity at time t, $\tau_1$ and $\tau_2$ are the fast and slow decay components, respectively, $A_1$ and $A_2$ are constants [20]. At present, there is no clear explanation for the origin of the fast decay component ($\tau_2$). One possible explanation is the presence of $Cr^{3+}$ ions in dodecahedral sites [34]. Depending on the investigated samples, the luminescence decay curves of $Cr^{3+}$-doped garnet single crystals were monoexponentially [1,14], close to single exponential with slight deviation [2], or two-exponential decay [34,35]. Fig. 11 presents the luminescence lifetime temperature maps. The measured luminescence lifetimes $\tau_1$, and $\tau_2$ vary from 6.5 ms to 0.01 ms and 1.5 ms to 0.01 ms, respectively.

Increasing temperature and $Cr^{3+}$ concentration significantly reduce the lifetimes of both decay components, with drastic reductions observed in certain samples. Both increasing temperature and $Cr^{3+}$ concentration ions lead to a decrease in the measured lifetimes of slow ($\tau_1$) and fast ($\tau_2$) components. The calculated $\tau_1$ values for Cr:YGG nanocrystals at -190 °C were 6,5 ms (Cr0.1), 5,1 ms (Cr0.5), 5,0 ms (Cr1), 1,6 ms (Cr3), and 0,3 ms (Cr5). The calculated $\tau_2$ at -190 °C were 1,5 ms (Cr0.1), 0,9 ms (Cr0.5), 0,8 ms (Cr1), 0,2 ms (Cr3), and 0,05 ms (Cr5). A temperature increase results in a drastic reduction in both $\tau_1$ and $\tau_2$, by more than two orders of magnitude. For example, when the temperature increases from -190 °C to 500 °C in Cr0.1, $\tau_1$, and $\tau_2$ decrease from 6.5 ms to 0.01 ms and 1.5 ms to 0.01 ms, respectively. The calculated lifetimes are summarized in Table S2.

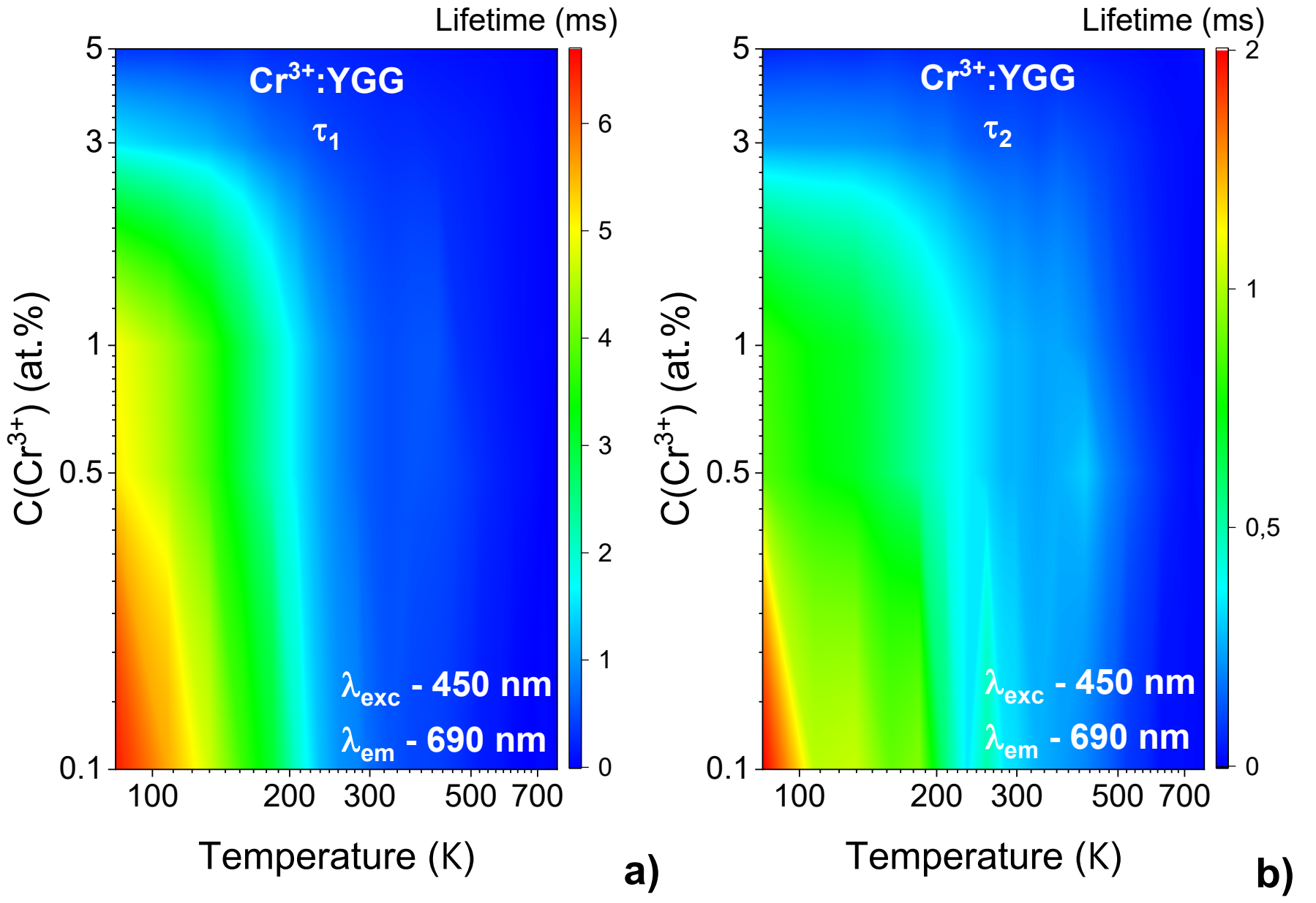

Fig. 11: Luminescence lifetime map of $Cr^{3+}$:YGG nanocrystals measured at $\lambda_{exc}$ – 450 nm.

The measured trends in the luminescence lifetimes do not align with the proposed earlier multicenter nature of $Cr^{3+}$ luminescence. The measured excitation and emission spectra indicate the presence of at least two $(CrO_6)^{9-}$ optically active centers, distinguished by a difference in crystal field strength. These $Cr^{3+}$ ions exhibit different deexcitation probabilities from their respective emission levels. The increase in emission intensity for the Cr3, and Cr5 samples in -100 °C to 100 °C range suggests a redistribution of absorbed energy in favor of $(CrO_6)^{9-}$ centers with higher crystal field strength. A similar trend would be expected in the measured lifetime; however, this was not observed in $Cr^{3+}$:YGG nanopowders (Figs. 4,7).

The energy transfer between the lowest excitation level of $Cr^{3+}$ions and between ions themselves complicates the lifetime trend analysis, causing difficulties in the proposed explanation for the absence of the expected lifetime trend. The measured decay curves of $Cr^{3+}$ ions (Fig. S5) depend on the depopulation rate of the energy levels from which emission occurs [7]. These decay curves were recorded at 690 nm, corresponding to the $R_1$-line transition of $Cr^{3+}$ ions at low temperatures. As temperature increases, energy transfer between the $^2E_g$ and $^4T_{2g}$ levels of $Cr^{3+}$ ions also increases, leading to an overall enhancement of the emission spectrum, particularly at 690 nm. This complicates the analysis of luminescence lifetimes in our samples. Additionally, energy transfer between $Cr^{3+}$ further increases the complexity of the system. Therefore, no clear explanation can currently be proposed for the absence of a distinct luminescence lifetime pattern, which contradicts the trends observed in the excitation (Fig. 4) and emission spectra (Fig. 7).

## 4. Discussion

$Cr^{3+}$ ions are sensitive to changes in the local crystal structure and composition, leading to varying spectroscopic properties between different phosphors. One of the key features of $Cr^{3+}$ ions as part of the Transition Metal (TM) family is their sensitivity to changes in the local crystal structure and composition. This sensitivity arises from the fact that the outer electrons of these ions are unshielded, resulting in a strong ion–lattice coupling in TM ions [7]. Consequently, various $Cr^{3+}$-doped phosphors exhibit changes in their spectroscopic properties depending on their composition. For example, the room temperature emission spectra gradually transition from a sharp, narrow red emission in $Cr^{3+}$:$Al_2O_3$ to a broadband NIR emission for $Cr^{3+}$:$ZnWO_4$. Among the many factors influencing their properties, the most significant are symmetry and the distance between the $Cr^{3+}$ and its surrounding ligands. A change in their chemical composition alters these parameters, modifying the emission spectra and potentially causing a redshift or even disappearance of the emission band. This behavior can be attributed to the electronic structure of the $Cr^{3+}$ ions.

Variation in the crystal field influences the positions of the lowest $Cr^{3+}$ energy levels, causing changes in the absorption and luminescence properties. The electronic structure of free $Cr^{3+}$ ions consists of three main energy levels: the ground state $^4F$, and the excited states $^4P$ and $^2G$. When $Cr^{3+}$ ions interact with the surrounding ligands, these energy levels undergo crystal field splitting, which ultimately shapes their optical properties. Among the resulting electronic states, the most relevant for optical properties are the $^4A_{2g}(^4F)$ ground state and the $^4T_{2g}(^4F)$, $^4T_{1g}(^4F)$ excited states (originated from the $^4F$ level), and $^2E_g$ excited state (originated from $^2G$ level). Due to the same electron–lattice coupling behavior of the $^4A_{2g}(^4F)$ ground state and the $^2E_g(^2G)$ excited state, transitions between these levels result in sharp and narrow and sharp optical features, so-called R-lines, which are observed in both absorption and emission spectra [7]. This $^4A_{2g}(^4F)\leftrightarrow{}^2E_g(^2G)$ transition is almost independent of crystal field strength and therefore appears at nearly the same position across different host lattices. In contrast, $^4A_{2g}(^4F)\leftrightarrow{}^4T_{2g}(^4F)$ transitions are strongly dependent on the crystal field strength. As a result, the corresponding optical absorption and emission bands can vary significantly in positions depending on the specific octahedral environment of the $Cr^{3+}$ ions [7]. In other words, these crystal-field-sensitive transitions change noticeably from one host crystal to another, making them useful probes of the local structure around $Cr^{3+}$ ions.

Influence of the crystal field on the $Cr^{3+}$ luminescence can be described by the Racah parameters and crystal field splitting parameter Dq, where Dq/B ranges from weak, intermediate, to strong. The influence of the host lattice on the spectral positions of $Cr^{3+}$ energy levels can be described using the Racah parameters (A, B, and C) and the crystal field splitting parameter Dq. The variation in the energy levels of $Cr^{3+}$ ions occupying an octahedral site as a function of crystal field strength (Dq/B) is represented in the Sugano–Tanabe diagrams. Broadly, the spectrum of Dq/B values can be classified into three categories: weak, intermediate, and strong crystal field. These categories are determined by the position at which the $^2E_g$ and $^4T_{2g}$ energy levels intersect in the diagram.

At low temperatures, $Cr^{3+}$ ions in strong crystal fields emit sharp R-lines, while those in weak fields produce broad emission bands, and intermediate ones show both types of emission:

- $Cr^{3+}$ ions in a strong crystal field exhibit emission originating only from $^2E_g\rightarrow{}^4A_{2g}$ electronic transition, resulting in sharp R-lines.
- $Cr^{3+}$ ions in a weak crystal exhibit emission that originates from the $^4T_{2g}\rightarrow{}^4A_{2g}$ electron transition, producing a broad emission band.
- $Cr^{3+}$ ions in intermediate crystal field exhibit both R-lines and broadband emission, due to the contribution of both $^2E_g$ and $^4T_{2g}$ emission levels.

The increase in temperature leads to the appearance of broadband emission in $Cr^{3+}$ ions in a strong crystal field, while only broadband emission is detected for $Cr^{3+}$ in a weak field. Due to the spin-forbidden nature of the $^4A_{2g} \leftrightarrow {}^2E_g$ transition (contrast to the spin-allowed $^4A_{2g} \leftrightarrow {}^4T_{2g}, {}^4T_{1g}$), the lifetime of the $^2E_g$ state is significantly longer compared to the $^4T_{2g}$ state [7]. Thus, in a strong crystal field, where $^2E_g$ is the lowest excited state, thermalization can populate the $^4T_{2g}$ level from the $^2E_g$ level, resulting in both $^2E_g \rightarrow {}^4A_{2g}$ R-lines and $^4T_{2g} \rightarrow {}^4A_{2g}$ emission components at elevated temperatures. Consequently, when $^4T_{2g}$ is the lowest excited state (i.e. in a weak crystal field), only broadband emission is observed across the entire temperature range.

The presence of R-lines in $Cr^{3+}$ luminescence spectra is a clear sign of the emission from ions in a strong crystal field. The presence of R-lines at -190 °C in all samples indicates that the emission originates from $Cr^{3+}_{STRONG}$. It has been noted that the position of R-lines is nearly independent of the host's crystal structure. However, a slight shift in the R-line maxima can still occur in different phosphor materials. For example, previous studies have shown that changing the host composition from $Y_3Al_5O_{12}$ to $Lu_3Al_5O_{12}$ causes a blueshift of R-lies approximately at 50 $cm^{-1}$ [36]. Similarly, the $R_1$-line position of $Cr^{3+}$:LuSGG (Dq/B - 2.57) and YSGG (Dq/B - 2.50) were 14513 $cm^{-1}$ and 14450 $cm^{-1}$, respectively [2].

The small difference in $R_1$-line positions in all samples suggests similar local crystal fields around the ions; however, low-temperature spectra suggest the presence of a small concentration of $Cr^{3+}$ ions in a low crystal field. In our study, the difference in $R_1$-line position between samples was smaller than 5 $cm^{-1}$, suggesting (though not definitely proving) a similar local crystal field around $Cr^{3+}_{STRONG}$ ions in all samples. R-lines position is connected to the change of C Racah parameter, while the B parameters do not have a clear influence on the position of the $^2E_g$ state [2]. The similarity in $^4A_{2g} \rightarrow {}^4T_{2g}$ excitation maximums (Fig. 5(d)), $^2E_g$ energy levels positions (Fig. 8(f), and the calculated Racah parameters (Table 2) support the conclusions that the comparable emission spectra should be expected across all samples, as is indeed observed at higher temperatures. However, this consistency does not hold for the low-temperature spectra, suggesting that an additional emission center might be present in the samples.

At low temperatures, the R-lines emission and NIR centered broadband emission originate from different $Cr^{3+}$ ions. The measured low-temperature luminescence spectra reveal the presence of an NIR emission band in samples with $Cr^{3+}$ concentration of 3 at.% and larger. This band originates from the $^4T_{2g} \rightarrow {}^4A_{2g}$ electronic transitions, with emission maxima observed at 775 nm for Cr3, 800 nm for Cr5, and 875 nm for Cr10. It should be noted that the luminescence spectra of Cr:GSGG single crystal at 4 K also show a broadband emission with an emission maximum around ~750 nm [14]. Since Cr:GSGG is considered to be in the intermediate crystal field region ($^4T_{2g}$ level is 50 cm$^{-1}$ below $^2E_g$ level), the $^4T_{2g}$→$^4A_{2g}$ broadband emission with $\lambda_{max}$ > 750 nm is attributed to $Cr^{3+}_{WEAK}$ ions. Based on this, the low-temperature luminescence spectra of the Cr3, Cr5, and Cr10 samples can be interpreted as arising from two types of $(CrO_6)^{9-}$ optically active centers:

- $Cr^{3+}_{STRONG}$ via sharp R-lines emission originating from the $^2E_g$→$^4A_{2g}$ transition.
- $Cr^{3+}_{WEAK}$ via broadband NIR emission originating from the $^4T_{2g}$→$^4A_{2g}$ transition.

Increased temperature causes a redistribution of the contribution to luminescence between $Cr^{3+}$ ions in the weak and strong crystal field. An increase in temperature leads to the disappearance of emissions from $Cr^{3+}_{WEAK}$ ions, leaving only the emission from $Cr^{3+}_{STRONG}$ ions (Fig. 7). This behavior can be attributed to the higher rate of nonradiative relaxation processes in $Cr^{3+}_{WEAK}$ ions, which results in a relative increase in contribution from $Cr^{3+}_{STRONG}$ ions at elevated temperatures. In $Cr^{3+}$-doped garnet single crystals, the measured luminescence quenching temperature $T_{1/2}$ has been shown to decrease with decreasing crystal field strength [2]. The difference in $T_{1/2}$ values for different $(CrO_6)^{9-}$centers may explain the observed emission intensity pattern in Cr5, and Cr10 samples (i.e., *decrease → unchanged → decrease*, Fig. 9). The recorded changes in emission intensity reflect the combined contribution from all $(CrO_6)^{9-}$ centers. The initial decline in emission intensity is attributed to the thermal quenching of $Cr^{3+}_{WEAK}$, while the emission of $Cr^{3+}_{STRONG}$ ions remain relatively stable up to a characteristic quenching temperature $T_c$ (Fig. 9). However, the proposed explanation does not account for the pattern observed in the R-lines intensities for Cr3 and Cr5 samples (Fig. 8(d)).

The redistribution in luminescence contribution occurs due to the $Cr^{3+}$→$Cr^{3+}$ energy transfer, which increases with the $Cr^{3+}$. One of the notable features of $Cr^{3+}$ ions is their ability to transfer energy, both among themselves and to other ions [19]. Efficient $Cr^{3+}$→$Yb^{3+}$ energy transfer has been demonstrated in Cr,Yb:YAG nanocrystals [20]. In our case, the decrease in the measured $Cr^{3+}$ luminescence lifetime with increasing concentration of $Cr^{3+}$ ions indicates that the $Cr^{3+}$→$Cr^{3+}$ energy transfer occurs even at low dopant levels Fig. 11, Fig. S6). The energy transfer between $Cr^{3+}_{STRONG}$ and $Cr^{3+}_{WEAK}$ ions affect the observed trends in spectroscopic behavior. The measured excitation spectra suggest that the emission from $Cr^{3+}_{WEAK}$ ions arises from $Cr^{3+}_{STRONG}$→$Cr^{3+}_{WEAK}$ energy transfer (Fig. 10). Furthermore, the observed changes in R-lines intensity (Fig. 8(d)) could be attributed to the variations in the efficiency of $Cr^{3+}$↔$Cr^{3+}$ energy transfer. As temperature increases, the possibility of radiative relaxation in $Cr^{3+}_{STRONG}$ ions may increase, reducing the likelihood that the absorbed energy is transferred to $Cr^{3+}_{WEAK}$ ions.

$Cr^{3+}$↔$Cr^{3+}$ energy transfer process might explain the change in the overall luminescence intensity in the samples. If the observed changes in spectroscopic parameters for the Cr3, Cr5, and Cr10 samples are influenced by energy transfer processes between $Cr^{3+}$ ions, then it is reasonable to assume that similar processes also affect the rest of the samples. In fact, $Cr^{3+}$↔$Cr^{3+}$ energy transfer likely occurs across all samples. The increase in overall $Cr^{3+}$ luminescence intensity for Cr0.1, Cr0.5, Cr1, and Cr3 samples (Fig. 9) could be attributed to the enhanced $Cr^{3+}$↔$Cr^{3+}$ energy transfer efficiency. This enhancement would lead to a redistribution of the absorbed energy, favoring relaxation through $Cr^{3+}_{STRONG}$ ions, which are characterized by a lower nonradiative transition rate. Additionally, the observed increase in luminescence intensity (Fig. 9) could also be partly due to a rise in the absorption strength with increasing temperature [2], further contributing to the emission enhancement.

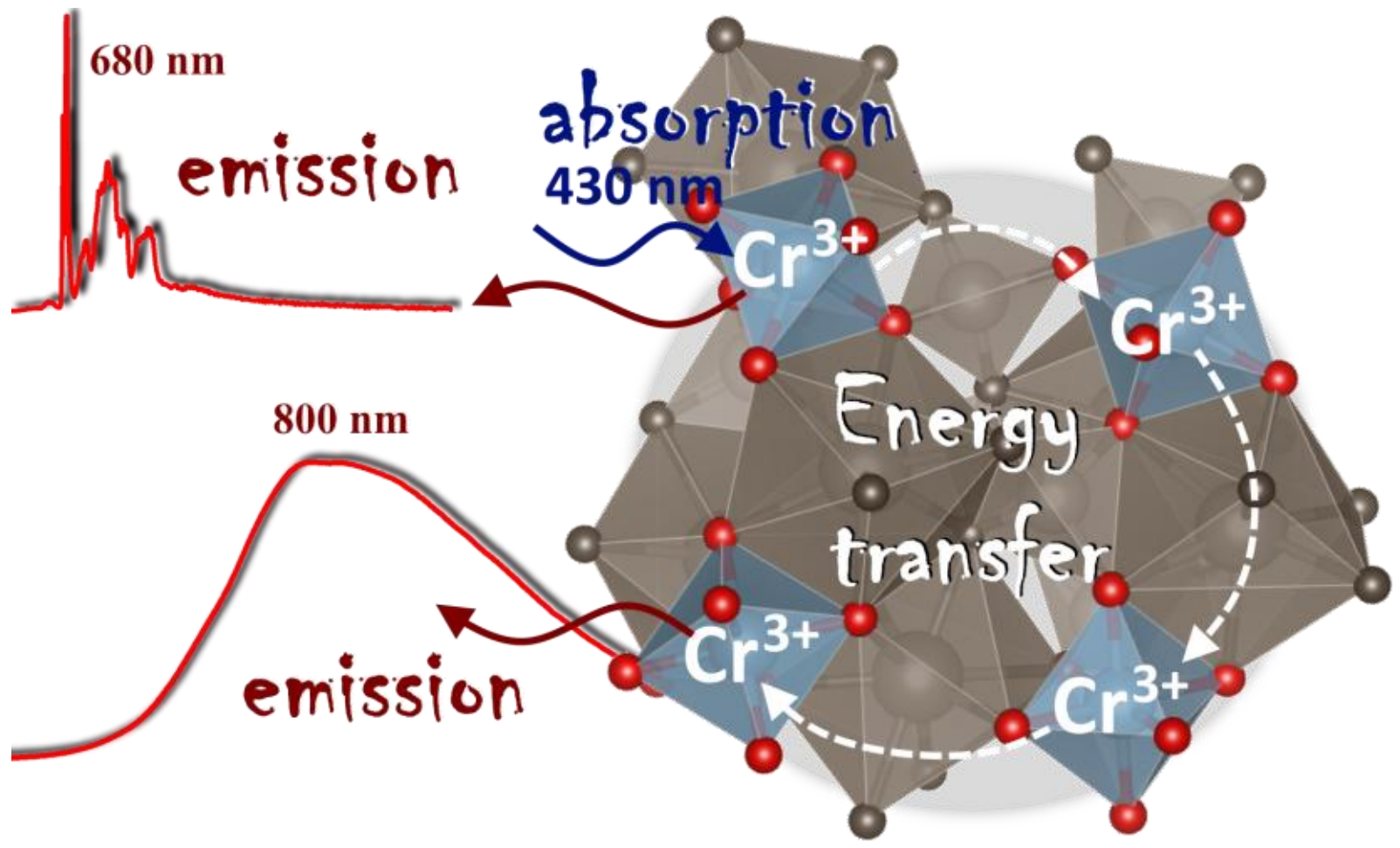

Fig. 12: schematic illustration of the energy transfer process between the $Cr^{3+}$ ions.

Probably, that energy transfer between the $Cr^{3+}$ ions occurs several times prior to emission, creating a chain of energy transfer. Non-uniform distribution of local crystal fields around $(CrO_6)^{9-}$ centers leads to the formation of multiple $Cr^{3+}$ emission centers with distinct optical properties. Energy transfer between these centers modifies the pattern of the observed luminescence spectra across different temperatures (Fig. 12). As a result, variations in luminescence spectra are observed even among $Cr^{3+}$-doped garnets with similar Racah and crystal field parameters. Due to efficient energy transfer, the energy of an absorbed photon can be relayed between multiple $Cr^{3+}$ ions before being emitted. Initially, one $Cr^{3+}$ ions absorb the photon and transfers the gained energy to another $Cr^{3+}$ ion, which in turn transfers to a third, and so one - until the energy is eventually emitted as luminescence by the final $Cr^{3+}$ ions in the sequence. Let us refer to this multi-step energy migration process as a “chain” of energy transfer.

The increase of $Cr^{3+}$ concentrations causes to increase in the length of this chain, while $Cr^{3+}$ ions in a low crystal field are likely to break this chain through the luminescence. The ”length” of this energy transfer chain increases with the $Cr^{3+}$ ions concentration, which in turn raises the likelihood of involving $Cr^{3+}_{WEAK}$ ions in the process. Due to their higher radiative transition rates compared to $Cr^{3+}_{STRONG}$ ions, $Cr^{3+}_{WEAK}$ ions are more likely to emit a photon than to further transfer the energy. This means that $Cr^{3+}_{WEAK}$ ions interrupt or “break” the energy transfer, releasing the absorbed energy via low-energy $^4T_{2g}\rightarrow{}^4A_{2g}$ transitions. Therefore, at higher $Cr^{3+}$ doping levels, the increased participation of $Cr^{3+}_{WEAK}$ ions results in the appearance of broadband NIR emission with $\lambda_{max} > 750$ nm, which is characteristic feature of $Cr^{3+}_{WEAK}$ ions emission.

Alternative explanations, like agglomeration or variations in crystal field strength, could be the reason for the detected patterns, but seem less significant than the chain energy transfer mechanism. Alternative explanations could also account for the observed patterns in the luminescence properties of $Cr^{3+}$ ions. For instance, the detected changes might be influenced by the agglomeration of $Cr^{3+}$ ions or variations in the crystal field strength parameter. These processes could indeed affect luminescence behavior $Cr^{3+}$:YGG nanophosphors. However, their impact appears to be less significant than the proposed ”chain” energy transfer phenomenon. It should be noted that current experimental data is not conclusive enough to definitively confirm this proposed mechanism. Therefore, theoretical modeling of the energy transfer and related processes within the samples would be highly valuable to validate and support this hypothesis.

The "chain" energy transfer mechanism could improve $Cr^{3+}$-doped phosphors, allowing for tunable emission across a wide range, but further studies are needed to confirm this hypothesis. The concept of a “chain” energy transfer mechanism offers a promising pathway for improving the functional characteristics of $Cr^{3+}$-doped phosphors. It suggests that by selecting appropriate host materials with low concentrations of luminescence quenching centers, it may be possible to engineer phosphors whose emission maximum can be tuned across a wide range, from visible to near infrared, depending on temperature. However, to fully explore and confirm this potential, further experimental studies should be conducted, particularly using $Cr^{3+}$-doped phosphors with similar chemical compositions.

**Conclusions**

Cr:YGG nanocrystals were synthesized using the Pechini method with the following compositions: $Y_3Ga_{(1-x)2}Cr_{2x}Ga_3O_{12}$, where x ranged from 0 to 0.7. An increase in the $Cr^{3+}$ concentration to 10 at.% and above leads to the formation of two impurity phases: $Y_2O_3$ and $Ga_2O_3$. The shape and broadening of the garnet peaks suggest the presence of two fractions of the garnet phase, with average crystalline sizes of approximately ~20 nm and ~50 nm, each exhibiting different lattice parameters. This observation was further supported by TEM analysis, which confirmed the presence of two distinct garnet fractions. The crystal field strength parameter Dq/B was found to increase from 2.14(5) to 2.32(5) with rising $Cr^{3+}$ concentration. The calculated Dq/B values for all samples were above the crossover point between $^2E_g$ and $^4T_{2g}$ levels, which lies between 0.20-0.24 Dq/B. This suggests that, based on purely crystal field considerations, the luminescence spectra of $Cr^{3+}$ ions should be similar across all samples.

The collected temperature-dependent excitation spectra of $Cr^{3+}$ ions revealed two distinct patterns: Cr01, Cr05, and Cr1 exhibit a decrease in excitation intensity with increasing temperature, while Cr03, Cr5, and Cr10 showed a maximum in excitation intensity at elevated temperatures. This behavior suggests that $Cr^{3+}$ concentration not only affects the morphology of the synthesized Cr:YGG nanocrystals but also influences their distribution between smaller and larger crystalline fractions. At 100 °C, the measured luminescence spectra of $Cr^{3+}$:YGG nanopowders displayed a similar spectral shape across all samples. However, low-temperature luminescence spectra exhibited a redshift in the emission maximum of the $^4T_{2g}\rightarrow{}^4A_{2g}$ broadband transition with increasing $Cr^{3+}$ concentration. The observed changes in both R-lines and broadband emission intensities can be attributed to the presence of two types of $(CrO_6)^{9-}$ optically active centers: $Cr^{3+}_{STRONG}$ and $Cr^{3+}_{WEAK}$ ions. The spectral overlap and energy transfer between these different $(CrO_6)^{9-}$ centers result in the detected changes in both $^4T_{2g}\rightarrow{}^4A_{2g}$ broadband emission and R-lines emission, depending on the $Cr^{3+}$ concentration.

The non-uniform distribution of local crystal fields around $(CrO_6)^{9-}$ units leads to the formation of multiple emission centers with distinct optical properties. Energy transfer between these optically active centers alters the pattern of the luminescence spectra observed at different temperatures. Due to the efficient $Cr^{3+}\rightarrow Cr^{3+}$ energy transfer, the energy absorbed by one $Cr^{3+}$ ion can be transferred along a sequence of $Cr^{3+}$ ions before being emitted, forming that can be described as a ”chain” of energy transfer. As the $Cr^{3+}$ concentration increases, the length of this chain increases as well, raising the probability of involving $Cr^{3+}$ ions located in a low crystal field environment. These ions tend to break the chain by emitting a photon via the $^4T_{2g}{\rightarrow}^4A_{2g}$ transition, resulting in NIR emission. The observed patterns in the luminescence behavior can potentially be explained by other mechanisms as well, such as $Cr^{3+}$ agglomerations or variation in Racah parameters. Therefore, additional experimental data are required to further validate or refute the proposed energy transfer chain model.

**Acknowledgement**

This work was supported by the Polish National Science Center, grant: OPUS 23, UMO-2022/45/B/ST5/01487. This work has been co-financed by the European Union under the HORIZON.1.2 – Marie Skłodowska-Curie Actions (MSCA), topic HORIZON-MSCA-2023-SE-01 – MSCA Staff Exchanges 2023, “$Cr^{4+}$:YAG/Polymer nanocomposite as alternative materials for Q-switched lasers: properties, modeling, and applications – ALTER-Q” - Project number 101182995.

**Literature**

[1] B. Malysa, A. Meijerink, T. Jüstel, Temperature dependent luminescence Cr3+-doped GdAl3(BO3)4 and YAl3(BO3)4, J Lumin 171 (2016) 246–253. https://doi.org/10.1016/J.JLUMIN.2015.10.042.

[2] B. Malysa, A. Meijerink, T. Jüstel, Temperature dependent Cr3+ photoluminescence in garnets of the type X3Sc2Ga3O12 (X = Lu, Y, Gd, La), J Lumin 202 (2018) 523–531. https://doi.org/10.1016/J.JLUMIN.2018.05.076.

[3] T.H. Maiman, Stimulated optical radiation in Ruby, Nature 187 (1960) 493–494. https://doi.org/10.1038/187493a0.

[4] M. Chaika, Advancements and challenges in sintering of $Cr^{4+}$:YAG: A review, J Eur Ceram Soc 44 (2024) 7432–7450. https://doi.org/10.1016/J.JEURCERAMSOC.2024.05.050.

[5] M.C.P.B. Joan Josep Carvajal Martí, ed., Luminescent Thermometry, 1st ed., Springer International Publishing, 2023. https://doi.org/10.1007/978-3-031-28516-5.

[6] P. Gluchowski, D. Hreniak, W. Lojkowski, W. Strek, Optical Properties of Cr(III) doped YAG Nanoceramics, ECS Trans 25 (2009) 113–119. https://doi.org/10.1149/1.3211168/XML.

[7] J.G. Solé, L.E. Bausá, D. Jaque, An Introduction to the Optical Spectroscopy of Inorganic Solids, John Wiley and Sons, 2005. https://doi.org/10.1002/0470016043.

[8] V. Stadnik, V. Hreb, A. Luchechko, Y. Zhydachevskyy, A. Suchocki, L. Vasylechko, Sol-Gel Combustion Synthesis, Crystal Structure and Luminescence of Cr3+ and Mn4+ Ions in Nanocrystalline SrAl4O7, Inorganics 2021, Vol. 9, Page 89 9 (2021) 89. https://doi.org/10.3390/INORGANICS9120089.

[9] P. Gluchowski, W. Strek, Luminescence and excitation spectra of Cr3+:MgAl 2O4 nanoceramics, Mater Chem Phys 140 (2013) 222–227. https://doi.org/10.1016/j.matchemphys.2013.03.025.

[10] P. Gluchowski, M. Chaika, Crystal-Field Strength Variations and Energy Transfer in $Cr^{3+}$-Doped GGG Transparent Nanoceramics, Journal of Physical Chemistry C 128 (2024) 9641–9651. https://doi.org/10.1021/ACS.JPCC.4C01658/ASSET/IMAGES/LARGE/JP4C01658_0011.JPEG.

[11] L. Vasylechko, V. Stadnik, V. Hreb, Y. Zhydachevskyy, A. Luchechko, V. Mykhaylyk, H. Kraus, A. Suchocki, Synthesis, Crystal Structure and Photoluminescent Properties of Red-Emitting CaAl4O7:Cr3+ Nanocrystalline Phosphor, Inorganics (Basel) 11 (2023) 205. https://doi.org/10.3390/INORGANICS11050205/S1.

[12] M.G. Brik, C.-G. Ma, Theoretical Spectroscopy of Transition Metal and Rare Earth Ions, 2019. https://doi.org/10.1201/9780429278754.

[13] S. Sugano, Y. Tanabe, H. Kamimura, Multiplets of Transition Metals Ions, Pure Appl. Phys. 33 (1970) 348.

[14] B. Struve, G. Huber, The effect of the crystal field strength on the optical spectra of Cr3+ in gallium garnet laser crystals, Applied Physics B Photophysics and Laser Chemistry 36 (1985) 195–201. https://doi.org/10.1007/BF00704574.

[15] R. Tomala, K. Grzeszkiewicz, D. Hreniak, W. Strek, Downconversion process in Yb3+doped GdAG nanocrystals, J Lumin 193 (2018) 70–72. https://doi.org/10.1016/j.jlumin.2017.06.038.

[16] R. Tomala, L. Marciniak, J. Li, Y. Pan, K. Lenczewska, W. Strek, D. Hreniak, Comprehensive study of photoluminescence and cathodoluminescence of YAG:Eu3+ nano- and microceramics, Opt Mater (Amst) 50 (2015) 59–64. https://doi.org/10.1016/J.OPTMAT.2015.06.042.

[17] M.A. Chaika, P. Dluzewski, K. Morawiec, A. Szczepanska, K. Jablonska, G. Mancardi, R. Tomala, D. Hreniak, W. Strek, N.A. Safronova, A.G. Doroshenko, S. V. Parkhomenko, O.M. Vovk, The role of Ca 2+ ions in the formation of high optical quality Cr 4+ ,Ca:YAG ceramics, J Eur Ceram Soc 39 (2019) 3344–3352. https://doi.org/10.1016/j.jeurceramsoc.2019.04.037.

[18] H. Suo, Y. Wang, X. Zhang, W. Zheng, Y. Guo, L. Li, P. Li, Y. Yang, Z. Wang, F. Wang, A broadband near-infrared nanoemitter powered by mechanical action, Matter 6 (2023) 2935–2949. https://doi.org/10.1016/j.matt.2023.06.009.

[19] M. Chaika, R. Tomala, O. Bezkrovnyi, W. Strek, The influence of nonradiative relaxation on laser induced white emission properties in Cr:YAG nanopowders, J Lumin 257 (2023) 119734. https://doi.org/10.1016/j.jlumin.2023.119734.

[20] M. Chaika, R. Tomala, O. Bezkrovnyi, W. Strek, Spectroscopic properties of Cr,Yb:YAG nanocrystals under intense NIR radiation, Mater Res Bull 163 (2023) 112201. https://doi.org/10.1016/j.materresbull.2023.112201.

[21] D. Wang, X. Zhang, X. Wang, Z. Leng, Q. Yang, W. Ji, H. Lin, F. Zeng, C. Li, Z. Su, Investigation of the Structural and Luminescent Properties and the Chromium Ion Valence of Li2CaGeO4 Crystals Doped with Cr4+ Ions, Crystals 2020, Vol. 10, Page 1019 10 (2020) 1019. https://doi.org/10.3390/CRYST10111019.

[22] V.B. Mykhaylyk, H. Kraus, Y. Zhydachevskyy, V. Tsiumra, A. Luchechko, A. Wagner, A. Suchocki, Multimodal Non-Contact Luminescence Thermometry with Cr-Doped Oxides, Sensors 2020, Vol. 20, Page 5259 20 (2020) 5259. https://doi.org/10.3390/S20185259.

[23] Z. Song, P.A. Tanner, Q. Liu, Host Dependency of Boundary between Strong and Weak Crystal Field Strength of Cr3+ Luminescence, Journal of Physical Chemistry Letters 15 (2024) 2319–2324. https://doi.org/10.1021/ACS.JPCLETT.4C00008/ASSET/IMAGES/LARGE/JZ4C00008_0004.JPEG.

[24] M. Chaika, R. Lisiecki, K. Lesniewska-Matys, O.M. Vovk, A new approach for measurement of Cr4+ concentration in Cr4+:YAG transparent materials: Some conceptual difficulties and possible solutions, Opt Mater (Amst) 126 (2022) 112126. https://doi.org/10.1016/j.optmat.2022.112126.

[25] A. Luchechko, V. Vasyltsiv, Y. Zhydachevskyy, M. Kushlyk, S. Ubizskii, A. Suchocki, Luminescence spectroscopy of Cr3+ ions in bulk single crystalline β-Ga2O3, J Phys D Appl Phys 53 (2020) 354001. https://doi.org/10.1088/1361-6463/AB8C7D.

[26] A. Szysiak, R. Tomala, H. Węglarz, J. Kajan, M. Słobodzian, M. Chaika, Toward more performant eye safe lasers: Effect of increasing sensitizer amount in Yb3 +,Er3+:YAG transparent ceramic on its spectral characteristics, J Eur Ceram Soc 45 (2025) 117365. https://doi.org/10.1016/J.JEURCERAMSOC.2025.117365.

[27] V.B. Mykhaylyk, H. Kraus, L.I. Bulyk, I. Lutsyuk, V. Hreb, L. Vasylechko, Y. Zhydachevskyy, A. Wagner, A. Suchocki, Al2O3 co-doped with Cr3+ and Mn4+, a dual-emitter probe for multimodal non-contact luminescence thermometry, Dalton Transactions 50 (2021) 14820–14831. https://doi.org/10.1039/D1DT02836G.

[28] P. Głuchowski, R. Pazik, D. Hreniak, W. Strek, Luminescence studies of Cr3+ doped MgAl2O4 nanocrystalline powders, Chem Phys 358 (2009) 52–56. https://doi.org/10.1016/j.chemphys.2008.12.018.

[29] P. Głuchowski, R. Pazik, D. Hreniak, W. Strek, Luminescence properties of Cr3+:Y3Al5O12 nanocrystals, J Lumin 129 (2009) 548–553. https://doi.org/10.1016/j.jlumin.2008.12.012.

[30] Y. Zhydachevskyy, V. Mykhaylyk, V. Stasiv, L.I. Bulyk, V. Hreb, I. Lutsyuk, A. Luchechko, S. Hayama, L. Vasylechko, A. Suchocki, Chemical Tuning, Pressure, and Temperature Behavior of Mn4+Photoluminescence in Ga2O3-Al2O3Alloys, Inorg Chem 61 (2022) 18135–18146. https://doi.org/10.1021/ACS.INORGCHEM.2C02807/ASSET/IMAGES/LARGE/IC2C02807_0015.JPEG.

[31] M. Chaika, K. Elzbieciak-Piecka, O. Vovk, L. Marciniak, New explanation for oxidation-induced Cr4+ formation in garnets, Optical Materials: X 23 (2024) 100342. https://doi.org/10.1016/J.OMX.2024.100342.

[32] A. Bala, S. Rani, Down conversation visible emission spectra of Cr3+ doped gadolinium aluminium garnet, Opt Quantum Electron 55 (2023) 1–19. https://doi.org/10.1007/S11082-023-05105-Z/TABLES/4.

[33] A. Luchechko, V. Vasyltsiv, V. Stasiv, M. Kushlyk, L. Kostyk, D. Włodarczyk, Y. Zhydachevskyy, Luminescence spectroscopy of Cr3+ ions in bulk single crystalline β-Ga2O3-In2O3 solid solutions, Opt Mater (Amst) 151 (2024) 115323. https://doi.org/10.1016/J.OPTMAT.2024.115323.

[34] M. Yamaga, B. Henderson, K.P. O'Donnell, G. Yue, Temperature dependence of the lifetime of Cr3+ luminescence in garnet crystals, Applied Physics B 1990 51:2 51 (1990) 132–136. https://doi.org/10.1007/BF00326013.

[35] P.J. Dereń, A. Watras, A. Gągor, R. Pązik, Weak crystal field in yttrium gallium garnet (YGG) submicrocrystals doped with Cr 3+, Cryst Growth Des 12 (2012) 4752–4757. https://doi.org/10.1021/CG300435T/SUPPL_FILE/CG300435T_SI_004.PDF.

[36] L. Seijo, S.P. Feofilov, Z. Barandiarán, Fine-Tuning the Cr3+ R1-Line by Controlling Pauli Antisymmetry Strength, Journal of Physical Chemistry Letters 10 (2019) 3176–3180. https://doi.org/10.1021/acs.jpclett.9b00858.